\documentclass[aps,preprint,noshowpacs,superscriptaddress,groupedaddress]{revtex4}

\usepackage{dcolumn}
\usepackage[normalem]{ulem}
\usepackage{subfiles}
\usepackage{graphpap}
\usepackage[dvips]{graphicx}
\usepackage[dvips]{graphics}
\usepackage{color}
\usepackage{hyperref}
\usepackage{amsmath}
\usepackage{float}

\begin{document}
\title{Cooling and Self-Oscillation in a Nanotube Electro-Mechanical Resonator}

\author{C. Urgell}
\altaffiliation{These authors contributed equally to this work.}
\affiliation{ICFO - Institut De Ciencies Fotoniques, The Barcelona Institute of Science and Technology,
Mediterranean Technology Park, 08860 Castelldefels (Barcelona), Spain}
\author{W. Yang}
\altaffiliation{These authors contributed equally to this work.}
\affiliation{ICFO - Institut De Ciencies Fotoniques, The Barcelona Institute of Science and Technology,
Mediterranean Technology Park, 08860 Castelldefels (Barcelona), Spain}
\author{S. L. de Bonis}
\affiliation{ICFO - Institut De Ciencies Fotoniques, The Barcelona Institute of Science and Technology,
Mediterranean Technology Park, 08860 Castelldefels (Barcelona), Spain}
\author{C. Samanta}
\affiliation{ICFO - Institut De Ciencies Fotoniques, The Barcelona Institute of Science and Technology,
Mediterranean Technology Park, 08860 Castelldefels (Barcelona), Spain}
\author{M. J. Esplandiu}
\affiliation{Catalan Institute of Nanoscience and Nanotechnology (ICN2), CSIC and BIST, Campus UAB, Bellaterra, 08193 Barcelona, Spain}
\author{Q. Dong}
\affiliation{Centre de Nanosciences et de Nanotechnologies, CNRS, Univ. Paris-Sud, Univ. Paris-Saclay, C2N – Marcoussis, 91460 Marcoussis, France}
\author{Y. Jin}
\affiliation{Centre de Nanosciences et de Nanotechnologies, CNRS, Univ. Paris-Sud, Univ. Paris-Saclay, C2N – Marcoussis, 91460 Marcoussis, France}
\author{A. Bachtold}
\affiliation{ICFO - Institut De Ciencies Fotoniques, The Barcelona Institute of Science and Technology,
Mediterranean Technology Park, 08860 Castelldefels (Barcelona), Spain}


\begin{abstract}

{\bf
Nanomechanical resonators are used with great success to couple mechanical motion to other degrees of freedom, such as photons, spins, and electrons~\cite{treutlein2014,Aspelmeyer2014a}. Mechanical vibrations can be efficiently cooled and amplified using photons, but not with other degrees of freedom. Here, we demonstrate a simple yet powerful method for cooling, amplification, and self-oscillation using electrons. This is achieved by applying a constant (DC) current of electrons through a suspended nanotube in a dilution fridge. We demonstrate cooling down to $4.6\pm 2.0$ quanta of vibrations. We also observe self-oscillation, which can lead to prominent instabilities in the electron transport through the nanotube. We attribute the origin of the observed cooling and self-oscillation to an electrothermal effect. This work shows that electrons may become a useful resource for quantum manipulation of mechanical resonators.}
\end{abstract}


\maketitle
The vibrations of mechanical resonators have been coupled to electrons in different transport regimes, such as single-electron tunneling~\cite{Knobel2003,Woodside2002,Lassagne2009,Steele2009,Benyamini2014,Ares2016,Okazaki2016}, Kondo~\cite{Goetz2018}, and quantum Hall effect~\cite{Singh2012,Chen2015}. Because mechanical resonators are excellent force sensors, a small electrostatic force created by electrons generates a large displacement of the resonator. Conversely, the displacement reacts back on the electrons by a sizeable amount. This backaction of electrons on the resonator has been frequently studied by measuring the change in resonance frequency and in energy decay rate of vibrations~\cite{Knobel2003,Woodside2002,Lassagne2009,Steele2009,Benyamini2014,Ares2016,Okazaki2016,Goetz2018,Singh2012,Chen2015}. In principle, the backaction of electrons can also be used to cool and amplify thermal vibrational fluctuations and to generate self-oscillation by applying a DC electron current~\cite{Clerk2005,Armour2004}. Signatures of a modest cooling down to $\sim200$ quanta as well as self-oscillation were observed over a decade ago in a pioneer work~\cite{Naik2006} where a resonator is coupled to a superconducting single-electron transistor, but a quantitative understanding of the backaction has not been reported yet. Meanwhile, many theoretical schemes have been proposed to cool mechanical vibrations using electrons in different electron transport regimes; see for instance Refs.~\cite{Zippilli2009,Santandrea2011,Stadler2014,Arrachea2014a,Stadler2016}. However, these cooling schemes could not be implemented due to experimental difficulties. In this Letter, we show efficient backaction cooling in a current-biased suspended nanotube precooled in a dilution fridge.

Carbon nanotubes are a versatile system for the study of both electron transport and nanomechanics. Different electron transport regimes can be reached by tuning the transmission of electrons between the nanotube and electrodes~\cite{Laird2015}. Interaction can lead to electron attraction, Kondo behaviours, and Wigner states~\cite{Laird2015,Hamo2016,Deshpande2008}. On the other hand, carbon nanotubes are so small that they make the lightest mechanical resonators fabricated thus far. Cooling a nanotube resonator in a dilution fridge leads to high quality factors~\cite{Moser2014,Huttel2009}. As a result, the force sensitivity of the resonator is record high~\cite{Bonis2018}, and the effect of the electron-vibration coupling is expected to be especially large.

Suspending a carbon nanotube between two metal electrodes is key to form a nanomechanical resonator and to carry out  state-of-the-art electron transport measurements. This suppresses the electron backscattering in the nanotube due to the charge impurities and the rugosity of the substrate. We grow the carbon nanotube between two metal electrodes in the last step of the fabrication process using chemical vapour deposition in order to minimise residual contamination~\cite{Moser2014}. Measurements are carried out by applying a DC voltage to the source electrode ($V_\mathrm{sd}$) and the gate electrode ($V_\mathrm{g}$) patterned at the bottom of the trench (Fig.~1a). We detect the electrical current from the drain electrode using a RLC resonator with frequency $\omega_\mathrm{RLC}=2\pi\cdot 1.27$~MHz and a high-electron-mobility-transistor amplifier~\cite{Bonis2018}. We record the differential conductance $G_\mathrm{diff}$ of the device by applying an oscillating voltage $V_\mathrm{sd}^\mathrm{ac}$ to the source electrode with the frequency set at $\omega_\mathrm{RLC}$. Using a capacitive transduction scheme~\cite{Bonis2018}, we measure thermal vibrations with resonance frequency $\omega_\mathrm{0}$ by applying $V_\mathrm{sd}^\mathrm{ac}$ at the frequency $\omega_\mathrm{0}-\omega_\mathrm{RLC}$. In order to avoid perturbations from the measurement, we keep the amplitude of $V_\mathrm{sd}^\mathrm{ac}$ much smaller than $k_\mathrm{B}T/e$, where $k_\mathrm{B}$ is the Boltzmann constant, $T$ the temperature of the cryostat, and $e$ the electron charge. All the measurements presented here are carried out at the base temperature of the fridge except when stated differently.

Electron transport measurements indicate that electrons are in the Kondo regime~\cite{Laird2015}. A regular shell filling with Kondo ridges at zero source-drain bias is observed upon sweeping $V_\mathrm{g}$ (Figs.~1b,c). Unlike normal Coulomb blockade, $G_\mathrm{diff}$ increases in every second conductance valleys when decreasing temperature. This shows the SU(2) nature of the Kondo effect in this device.

Energy decay measurements of thermal vibrations reveal that the quality factor $Q=6.8\cdot10^6$ is remarkably high when compared to previous works~\cite{Moser2014,Huttel2009,Bonis2018}. This is also higher than the quality factor inferred from the spectral resonance linewidth, since the energy decay rate is smaller than the spectral resonance linewidth (Figs.~1d-f). The difference is attributed to dephasing. The resonance frequency can be tuned by sweeping both $V_\mathrm{g}$ and $V_\mathrm{sd}$ (Figs.~1g,h). The slopes $\partial \omega_0/  \partial V_\mathrm{g}$ and $-\partial \omega_0/  \partial V_\mathrm{sd}$ are often rather similar, suggesting that they are related to the same origin, that is, the mechanical tension induced by the static displacement of the nanotube. Thermal vibrations can be cooled with the cryostat down to $\sim 70$~mK (Fig.~1i). We attribute the saturation of the displacement variance at low temperature to the electric noise in the circuit.

We observe instabilities in the conductance arising periodically in $V_\mathrm{g}$ (see arrows in Fig.~1b). They emerge at finite source-drain bias in charge stability diagram measurements. In these instability regions, the peaks in conductance are truncated  (Figs.~2a-c) and conductance traces as a function of $V_\mathrm{g}$ appear noisy (Fig.~2d). While similar conductance instabilities were previously reported~\cite{Steele2009}, we will show below that these instabilities are related to large-amplitude vibrations.

The measured conductance instabilities originate from the switching of the mechanical motion between thermal noise and self-oscillation (Fig.~3). Upon sweeping $V_\mathrm{sd}$ through the instability region, the variance of the displacement  $\delta z^2$ dramatically increases (Fig.~3a), and the decay rate gets suppressed towards zero near the border of the instability region (shaded in yellow in Fig.~3b). These two experimental facts point to the development of self-oscillation. The phase-space and the histogram of the two quadratures of the motion can be described by the superposition of the distributions of a donut and a Gaussian-like peak (Figs.~3h,i), suggesting that the motion switches back and forth between self-oscillation with high $\delta z^2$ and thermal noise with low $\delta z^2$ (Sec.~III in Supplementary Information). This switching is further supported by the fact that the resonance lineshape is unusually broad (Fig.~3g); indeed, the large linewidth is then related to the large fluctuations of $\delta z^2$ and the resonator nonlinearity. Pure self-oscillation can also be observed without any switches to thermal vibrations; this often happens at higher $V_\mathrm{sd}$ values. The shift in resonance frequency due to electron backaction (Fig.~3c) is difficult to quantify, since the resonance frequency also depends on the mechanical tension induced by $V_\mathrm{sd}$, the temperature rise of the nanotube lattice due to Joule heating, and the variance of the displacement through the mechanical nonlinearity.

Mechanical vibrations are cooled down to $4.6 \pm 2.0$ quanta at $V_\mathrm{g}=-943$~mV upon increasing the source-drain bias to $V_\mathrm{sd}\simeq0.565$~mV (Figs.~4a-c). Cooling is accompanied with a strong increase of the decay rate (Fig. 4d), indicating a backaction effect. This efficient cooling occurs when the transconductance is negative and large (Fig.~4e). We observe cooling at other $V_\mathrm{g}$ values when the transconductance is negative as well (Sec.~II in Supplementary Information). The determination of the number of quanta is robust against the hypothetical miscalibration of the amplification chain and of the attenuation along the coaxial cables (Sec.~I in Supplementary Information). The uncertainty in the transconductance, which enters into the transduction of the displacement, is 5.7~\% of its value at $V_\mathrm{g}=-943$~mV and $V_\mathrm{sd}\simeq0.565$~mV. As explained above, the frequency shift due to backaction cannot be quantified, since the frequency shift depends on various other effects that are difficult to disentangle experimentally.

The nanotube experiences Joule heating due to the current flowing through the resonator. Backaction cooling predicts that the phonon occupation at finite bias is given by $n(V_\mathrm{sd})=\Gamma_\mathrm{bath}n_\mathrm{bath}/\Gamma_\mathrm{decay}(V_\mathrm{sd})$ where $n_\mathrm{bath}=k_\mathrm{B}T_\mathrm{bath}/\hbar \omega_\mathrm{0}$ is the thermal phonon number and $\Gamma_\mathrm{bath}$ is the coupling to the thermal bath. We would achieve much lower phonon occupation, if the bath temperature $T_\mathrm{bath}$ was given by the cryostat temperature, while setting $\Gamma_\mathrm{bath}$ to the measured decay rate at zero-bias. This indicates that Joule heating is sizable. We deduce the bath temperature from the measured values of $n$ and $\Gamma_\mathrm{decay}$ (Fig.~4f)~\cite{Song2014a,Weber2016}. The temperature rise can be well described by Joule heating for different $V_\mathrm{g}$ values using the phenomenological relation  $T_\mathrm{bath}=T_\mathrm{vib}^0+\eta G V_\mathrm{sd}^2$ (Fig.~4f), where $T_\mathrm{vib}^0$ is the measured vibration temperature at zero-bias, $G$ is the conductance, and $\eta$ is the same constant for all the $V_\mathrm{g}$ values. The temperature rise is not accounted for by the electrostatic force associated to the electron shot noise of the nanotube, since the temperature rise does not depend linearly on $V_\mathrm{sd}$ and it is independent of $V_\mathrm{g}$ to a first approximation. The shot noise of the suspended nanotube behaves in the usual way with a Fano factor between $0.2$ and $0.3$ (Sec.~IV of Supplementary Information).

Here, we discuss the possible origins of the observed backaction.  It could be related to the usual backaction in electro-mechanical resonators \cite{Knobel2003,Woodside2002,Lassagne2009,Steele2009,Benyamini2014,Ares2016,Okazaki2016,Goetz2018}, where conducting electrons generate an electrostatic force on the nanotube and the retardation of the force is given by the transmission of electrons between the nanotube and electrodes. However, we do not observe resonance frequency dips when sweeping $V_\mathrm{g}$ (Fig.~1g), showing that the strength of this backaction is weak. Moreover, this backaction predicts cooling at the conductance peaks~\cite{Clerk2005,Armour2004}, which is the opposite of what is observed in Fig.~2, that is, self-oscillation near conductance peaks. This shows that the backaction measured in electro-mechanical resonators at zero source-drain bias cannot describe our results at finite bias. Another possible mechanism could be related to the retardation created by the circuit, where the vibration-induced current noise of the nanotube generates a retarded electrostatic force due to the capacitance of the circuit. However, the predicted decay rate is too weak to produce the cooling observed in Figs.~4a,c. We conclude that backaction with electrostatic origins cannot account for our findings.

We attribute the origin of the backaction to an electrothermal effect~\cite{Steeneken2011}, which is an analogue of the photothermal backaction often observed in opto-mechanical resonators~\cite{Barton2012}. The power $GV_\mathrm{sd}^2$ of Joule heating modifies the mechanical tension in the nanotube through the effective thermal expansion coefficient of the device. This results in a net displacement $\delta z$ of the resonator when the nanotube is bent by e.g. the static electrostatic force associated with $V_\mathrm{g}$. This displacement reacts back on the dissipated power via $\delta G=\frac{dG}{dz}\delta z$ with a delay given by both the capacitance of the circuit and the thermalization time of the device~\cite{Steeneken2011}. This electrothermal effect modifies the decay rate by
\begin{equation}
\Delta \Gamma_\mathrm{back}=-\alpha\frac{dG_\mathrm{diff}}{dV_\mathrm{g}} \frac{C_\mathrm{g}^\prime}{C_\mathrm{g}} V_\mathrm{g} z_\mathrm{s} V_\mathrm{sd}^2.
\label{eq:electrothermalrate}
\end{equation}
Here, $C_\mathrm{g}$ is the capacitance between the nanotube and the gate electrode, $C_\mathrm{g}^\prime$ is its derivative with respect to $z$, and $z_\mathrm{s}$ is the static displacement. We use $\alpha$ as a free parameter in our analysis, since $\alpha$ depends on various quantities that are difficult to quantify. These include the thermalisation time, the effective thermal expansion coefficient, and the three-dimensional profile of the static bending of the nanotube.

The electron transport in the device controls the electrothermal backaction through $\frac{dG_\mathrm{diff}}{dV_\mathrm{g}}$. When $\frac{dG_\mathrm{diff}}{dV_\mathrm{g}}$ is positive and large, the total decay rate of the resonator can become effectively negative, leading to self-oscillation. When $\frac{dG_\mathrm{diff}}{dV_\mathrm{g}}$ is strongly negative so that $\Delta \Gamma_\mathrm{back}>>\Gamma_\mathrm{bath}$, the vibrations are efficiently cooled. Equation~\ref{eq:electrothermalrate} reproduces qualitatively the decay rate measured when increasing $V_\mathrm{sd}$ towards the self-oscillation regime shaded in yellow in Fig.~3b and to the strong cooling regime in Fig.~4d (see pink lines). See Sec. V of the Supplementary Information for more discussion on the different backactions.

This cooling method with electrons becomes efficient by precooling the resonator in a dilution fridge, so that the quality factor is high and the motion is sensitive to backaction. Electrothermal backaction is effective for small resonators, since the low mass increases the backaction strength, and the small heat capacity enhances the heating effect. Future studies may enable ground-state cooling. This may be achieved by enhancing the backaction rate (Eq.~\ref{eq:electrothermalrate}) using devices with higher transconductance and stronger coupling to the gate (to increase the $C_\mathrm{g}^\prime/C_\mathrm{g}$ ratio). Cooling mechanical vibrations with electrons may become a useful resource for quantum manipulation of mechanical resonators.


\textbf{Acknowledgments} This work is supported by the ERC advanced grant 692876, the Foundation Cellex, the CERCA Programme, AGAUR, Severo Ochoa (SEV-2015-0522), the grant FIS2015-69831-P of MINECO, and the Fondo Europeo de Desarrollo Regional (FEDER). We wish to thank Brian Thibeault for help in fabrication (UCSB).

{\textbf{Author contributions} W.Y. fabricated the devices with the support of C.U. and M.J.E. in the growth. C.U. and W.Y. carried out the measurements. C.U., W.Y., S.B., C.S., Q.D., and Y.J developed the detection circuit. C.U., W.Y., and A.B. did the analysis of the data and wrote the manuscript. A.B. supervised the work.

\begin{figure}[H]
	\label{characterization}
	\centering
	\includegraphics[scale=0.91, trim = 0cm 0cm 0cm 0cm, clip=true]{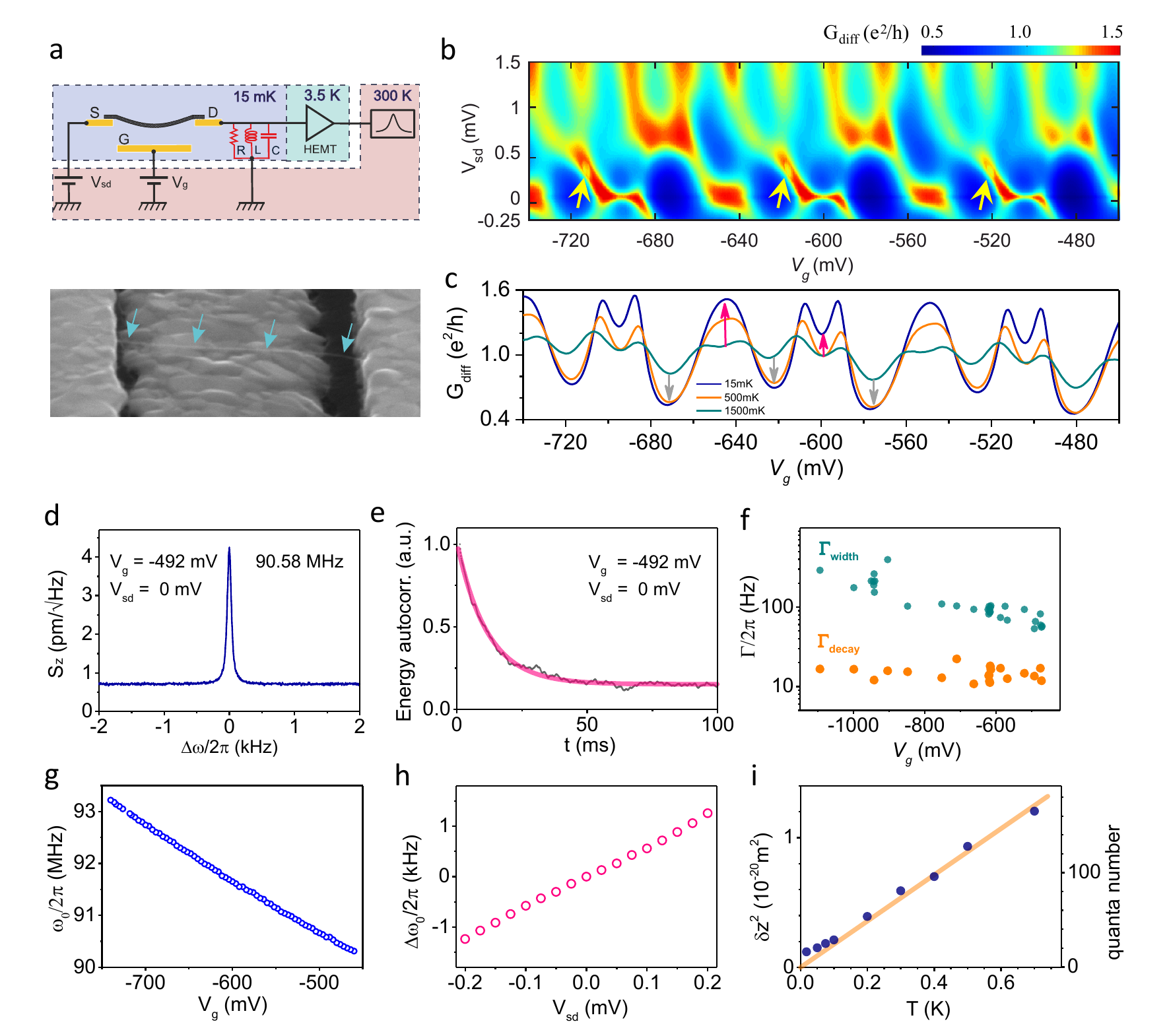}
	\caption{{\bf Characterization of the nanotube electro-mechanical resonator.} {\bf(a)} Measurement schematic and scanning electron microscopy image of the suspended nanotube. DC voltages $V_\mathrm{sd}$ and $V_\mathrm{g}$ are applied to electrodes $S$ and $G$, respectively. {\bf(b)} Differential conductance as a function of $V_\mathrm{sd}$ and $V_\mathrm{g}$. Yellow arrows indicate regions of conductance instabilities. {\bf(c)} Differential conductance as a function of $V_\mathrm{g}$ with $V_\mathrm{sd}=0$~mV for different temperatures. {\bf(d)} Displacement spectral density. {\bf(e)} Energy decay obtained from the autocorrelation of the time trace of $X^2+Y^2$, where $X$ and $Y$ are the two quadratures of thermal vibrations. The pink line indicates an exponential decay. {\bf(f)} Decay rate and spectral resonance width as a function of $V_\mathrm{g}$ with $V_\mathrm{sd}=0$~mV. {\bf(g,h)} Resonance frequency as a function of $V_\mathrm{g}$ and $V_\mathrm{sd}$. We set $V_\mathrm{sd}=0$~mV in {\bf g} and $V_\mathrm{g}=-560$~mV in {\bf h}. {\bf(i)} Variance of the displacement of thermal vibrations as a function of temperature at $V_\mathrm{g}=-185$~mV.}
\end{figure}

\begin{figure}[H]
	\label{instabilities}
	\centering
	\includegraphics[scale=0.95, trim = 0cm 0cm 0cm 0cm, clip=true]{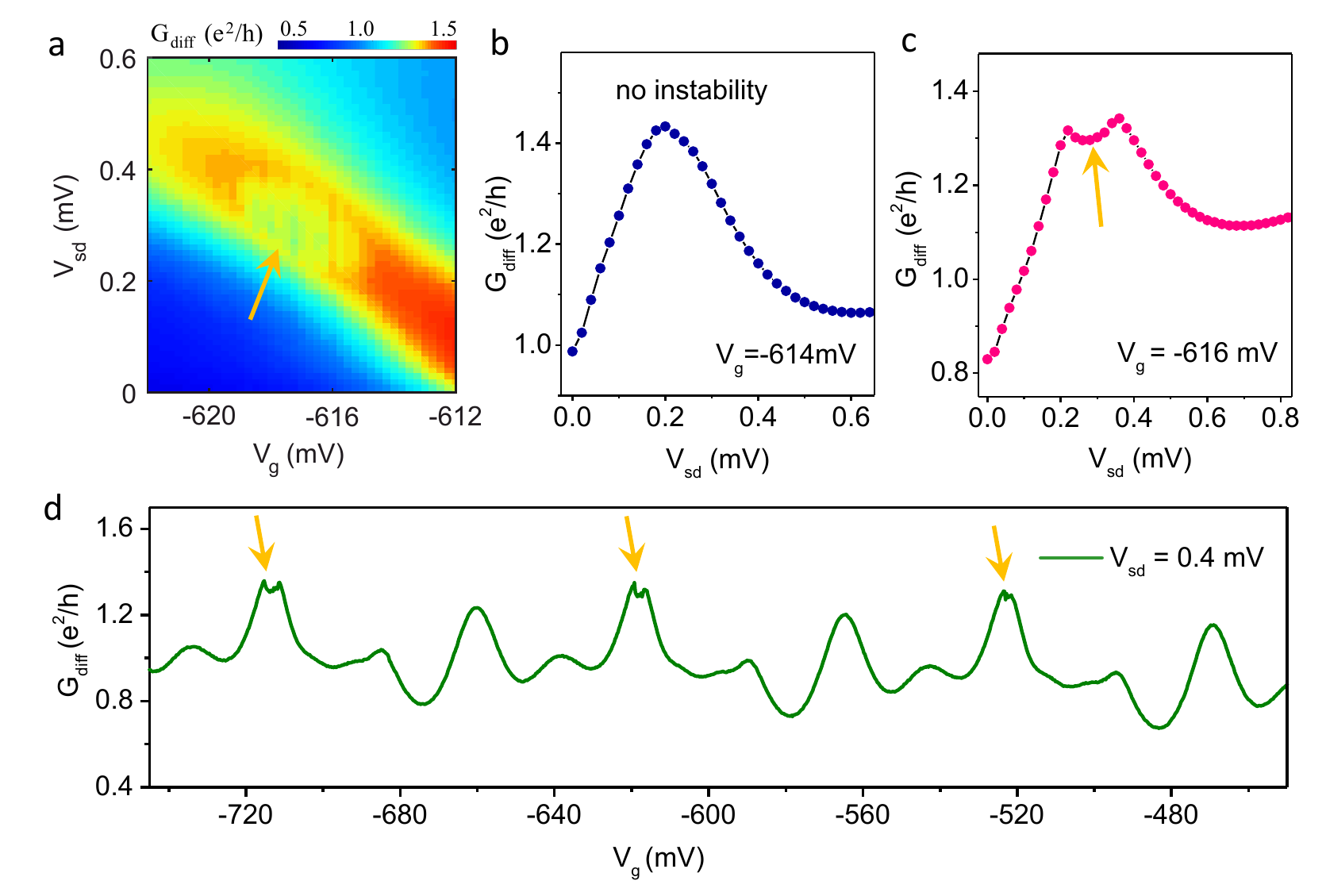}
	\caption{{\bf Conductance instabilities.} {\bf(a)} Differential conductance as a function of $V_\mathrm{sd}$ and $V_\mathrm{g}$. Yellow arrows in the different panels indicate regions of conductance instabilities. {\bf(b,c)} Differential conductance as a function of $V_\mathrm{sd}$ for two different $V_\mathrm{g}$ values. {\bf(d)} Differential conductance as a function of $V_\mathrm{g}$.
}
\end{figure}

\begin{figure}[H]
	\label{selfoscillation}
	\centering
	\includegraphics[scale=0.9, trim = 0cm 0cm 0cm 0cm, clip=true]{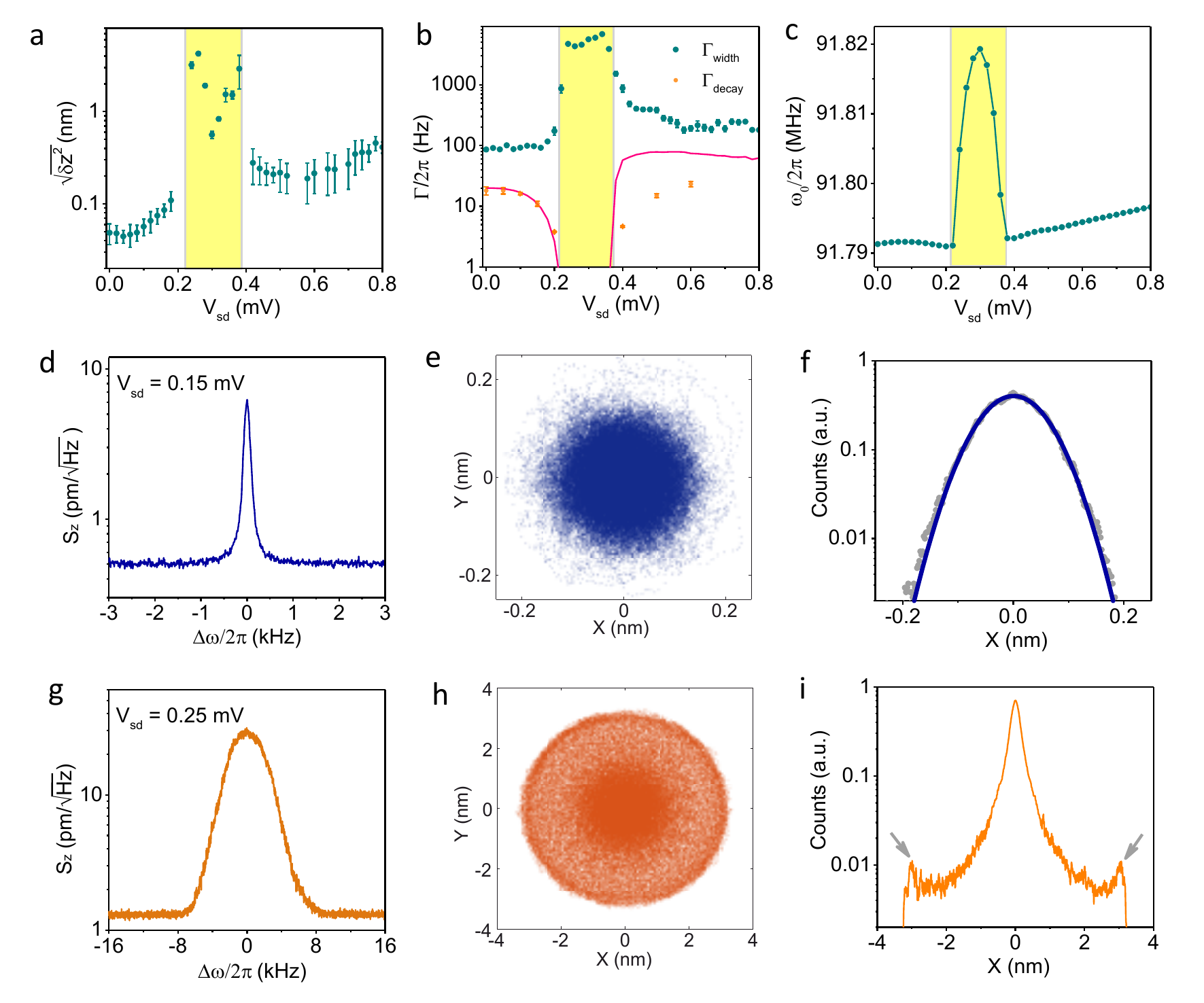}
	\caption{{\bf Self-oscillation at $V_\mathrm{g}=-616$~mV.} {\bf(a)} Variance of the displacement as a function of $V_\mathrm{sd}$. The yellow shaded area represents the region with self-oscillation.  {\bf(b)} Decay rate and spectral resonance width as a function of $V_\mathrm{sd}$. The pink line is the expected decay rate using Eq.~\ref{eq:electrothermalrate}. {\bf(c)} Resonance frequency as a function of $V_\mathrm{sd}$. {\bf(d-f)} Displacement spectral density, the phase-space of the two quadratures of the motion, and the associated histogram at $V_\mathrm{sd}=0.15$~mV. {\bf(g-i)} Same as {\bf d-f} but at $V_\mathrm{sd}=0.25$~mV. The arrows in {\bf(i)} indicate the donut distribution.}
\end{figure}

\begin{figure}[H]
	\label{cooling}
	\centering
	\includegraphics[scale=0.84, trim = 0cm 0cm 0cm 0cm, clip=true]{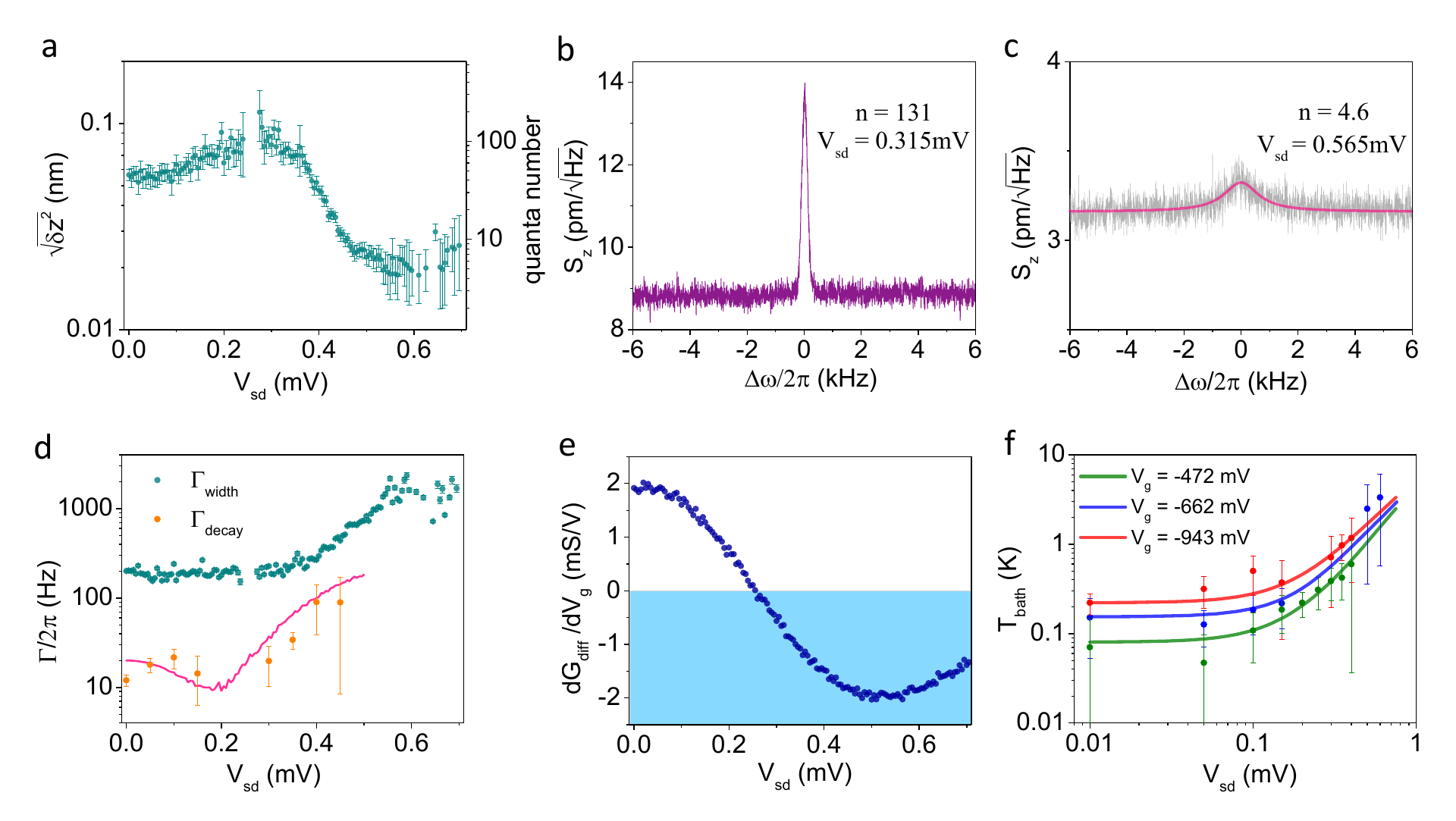}
	\caption{{\bf Cooling at $V_\mathrm{g}=-943$~mV.} {\bf(a)} Variance of the displacement as a function of $V_\mathrm{sd}$. The error bars arise from the uncertainty in the fitting of the resonance lineshape to a Lorentzian. {\bf(b,c)} Displacement spectral density at two different $V_\mathrm{sd}$. {\bf(d)} Decay rate and spectral resonance width as a function of $V_\mathrm{sd}$. The pink line is the expected decay rate using Eq.~\ref{eq:electrothermalrate}. {\bf(e)} Transconductance as a function of $V_\mathrm{sd}$. {\bf(f)} Bath temperature of mechanical vibrations as a function of $V_\mathrm{sd}$ for three different $V_\mathrm{g}$ values. The solid lines indicate the dependence expected from Joule heating; see text.}
\end{figure}
\newpage
\setcounter{figure}{0} 
\section{Detection of mechanical vibrations}\label{electrontransport}
Mechanical vibrations are electrically detected using a RLC resonator and a HEMT amplifier cooled at liquid-helium temperature (Fig.~1a of the main text)~\cite{Bonis2018}. Displacement modulation is transduced capacitively into current modulation by applying an input oscillating voltage $V_\mathrm{sd}^\mathrm{ac}$ across the nanotube. The frequency $\omega_\mathrm{sd}/2\pi$ of the oscillating voltage is set to match $\omega_\mathrm{sd}=\omega_\mathrm{0} \pm \omega_\mathrm{RLC}$, where $\omega_\mathrm{0}/2\pi$ is the resonance frequency of the nanotube resonator and $\omega_\mathrm{RLC}/2\pi$ the resonance frequency of the RLC resonator. Thermal vibrations are measured by recording the current noise at $\sim \omega_\mathrm{RLC}$. The current $\delta I$ is related to the displacement of the nanotube $\delta z$ by
\begin{eqnarray}
&& \delta I=\beta\delta z ,
\label{eq:Iz}\\
&& \beta=\frac{1}{2}\frac{dG_\mathrm{diff}}{dV_\mathrm{g}}V_\mathrm{g}V_\mathrm{sd}^\mathrm{ac}\frac{C_\mathrm{g}^{\prime}}{C_\mathrm{g}}.
\label{eq:beta}
\end{eqnarray}
Here, $\frac{dG_\mathrm{diff}}{dV_\mathrm{g}}$ is the transconductance, $V_\mathrm{g}$ is the static gate voltage, $C_\mathrm{g}$ is the capacitance between the nanotube and the gate electrode, and $C_\mathrm{g}^{\prime}$ is the derivative of $C_\mathrm{g}$ with respect to $z$. The spectral density $S_\mathrm{zz}$ of the displacement noise in the main text is obtained from the measured spectral density of the current noise using Eqs.~\ref{eq:Iz} and \ref{eq:beta}.

The calibration of the number of quanta is obtained in a reliable way thanks to the equipartition theorem
\begin{equation}
m\omega_0^2\delta z^2=k_\mathrm{B}T,
\label{eq:equipartition}
\end{equation}
where $m$ is the effective mass, $k_\mathrm{B}$ the Boltzmann constant, and $T$ the temperature. In practice, we measure the spectral density of the current noise to quantify the variance of the current $\delta I^2_\mathrm{res}$ associated to the mechanical resonance of thermal vibrations. The measurement of $\delta I^2_\mathrm{res}$ as a function of temperature in Fig.~1i of the main text determines $m\left(\frac{C_\mathrm{g}}{C_\mathrm{g}^{\prime}}\right)^2=2.5\times 10^{-33}\textrm{kg}\cdot \textrm{m}^2$
using Eqs.~\ref{eq:Iz}-\ref{eq:equipartition}. This allows us to quantify the effective temperature $T_\mathrm{vib}$ of the thermal vibrations at any $V_\mathrm{g}$ and DC voltage $V_\mathrm{sd}$ applied to the source electrode by measuring $\delta I^2_\mathrm{res}$  and $\frac{dG_\mathrm{diff}}{dV_\mathrm{g}}$ and using
\begin{equation}
T_\mathrm{vib}=m\left(\frac{C_\mathrm{g}}{C_\mathrm{g}^{\prime}}\right)^2\frac{4\omega_0^2}{k_\mathrm{B}\left(\frac{dG_\mathrm{diff}}{dV_\mathrm{g}}V_\mathrm{g}V_\mathrm{sd}^\mathrm{ac}\right)^2}\delta I^2_\mathrm{res}.
\label{eq:teffective}
\end{equation}
Importantly, the determination of $T_\mathrm{vib}$ does not depend on the hypothetical inaccurate calibration of the attenuation along the coaxial cables created by thermal contraction and of the amplification chain. Indeed, such inaccurate calibration, if sizeable, would have an effect on $V_\mathrm{sd}^\mathrm{ac}$ and $m\left(\frac{C_\mathrm{g}}{C_\mathrm{g}^{\prime}}\right)^2$, but it would be canceled out when determining $T_\mathrm{vib}$. The number of quanta of vibrations is obtained using $n=\frac{k_\mathrm{B}T_\mathrm{vib}}{\hbar \omega_0}-\frac{1}{2}$ with $\hbar$ the reduced Planck constant.

In order to quantify $S_\mathrm{zz}$ and $\delta z^2$, we estimate the capacitance $C_\mathrm{g}$ from the separation $\Delta V_\mathrm{g}=23.1$~mV between two
conductance peaks in the Coulomb blockade regime at large positive $V_\mathrm{g}$ values (Fig.~S1).
We obtain $C_\mathrm{g}=e/\Delta V_\mathrm{g}=6.94\times10^{-18}$~F. We get $C_\mathrm{g}^{\prime}=7\times10^{-12}\textrm{F/m}$ from the measurement of the variance of the displacement as a function of temperature using $m=2.7$~ag.

\begin{figure}[h!]
	\label{CB}
	\centering
	\includegraphics[scale=1, trim = 0cm 0cm 0cm 0cm, clip=true]{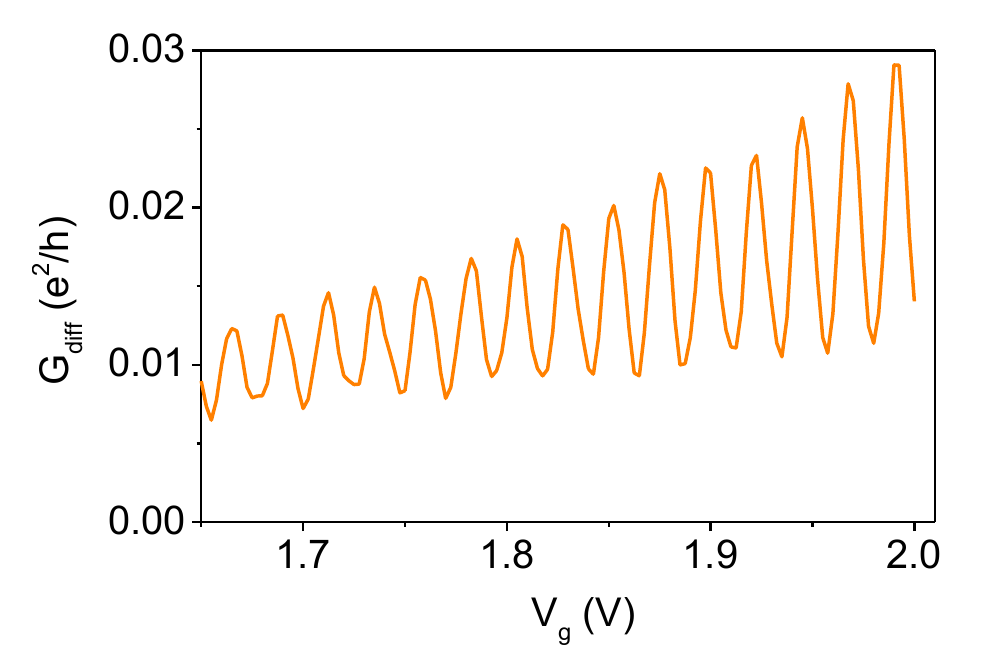}
	\caption{{\bf Coulomb blockade measurements.} Differential conductance $G_{diff}$ as a  function of gate voltage at 10~K and zero source-drain bias.}
\end{figure}

The energy decay rate $\Gamma_\mathrm{decay}$ is estimated by measuring the time trace of the two quadratures of thermal vibrations and by quantifying the autocorrelation of the amplitude squared. From these time trace measurements, we also obtain the phase-space of the two quadratures and the associated histogram (Fig.~3 of main text).

\begin{figure}[h!]
	\label{stabilitydiagram}
	\centering
	\includegraphics[scale=0.9, trim = 0cm 0cm 0cm 0cm, clip=true]{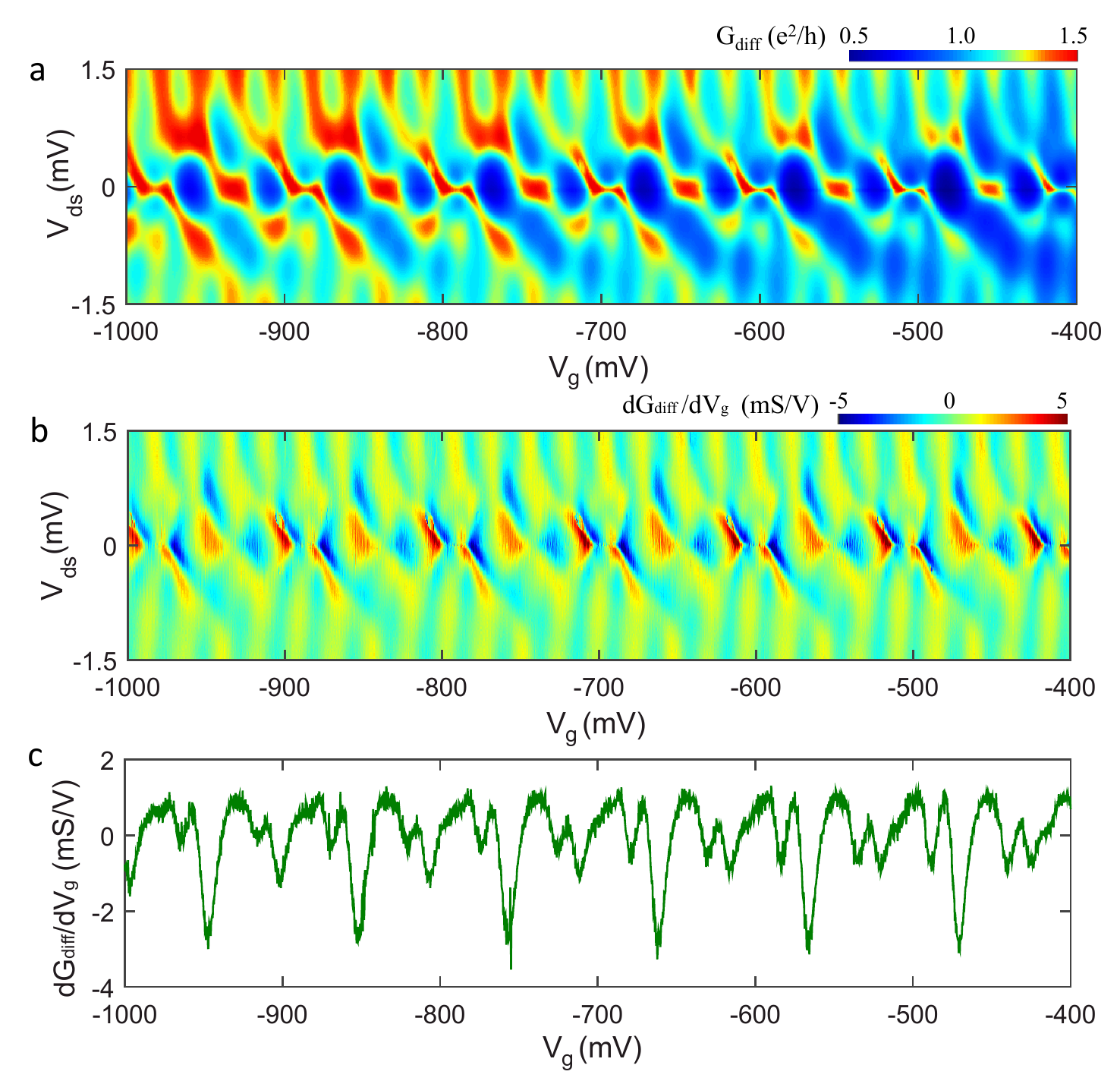}
	\caption{{\bf Electron transport measurements.} {\bf(a,b)} Differential conductance and transconductance as a function of $V_\mathrm{sd}$ and $V_\mathrm{g}$ measured at the base temperature of the fridge. {\bf(c)} Transconductance as a function of $V_\mathrm{g}$ measured at $V_\mathrm{sd}=0.7$~mV.}
\end{figure}

\section{Relation between electron transport and vibration cooling }

Figures~S2a,b show that the measurements of the differential conductance $G_{diff}$ and $\frac{dG_\mathrm{diff}}{dV_\mathrm{g}}$ as functions of $V_\mathrm{g}$ and $V_\mathrm{sd}$ are remarkably regular over a large range of gate voltage. This reflects the high quality of the nanotube. The shell filling with Kondo ridges at zero source-drain bias is observed over the full range of $V_\mathrm{g}$. The instability in the conductance discussed in Figs.~1b and 2d of the main text appears periodically in gate voltage over the full $V_\mathrm{g}$ range as well.

Figures~S2b,c show that regions with strongly negative $\frac{dG_\mathrm{diff}}{dV_\mathrm{g}}$ emerge periodically in $V_\mathrm{g}$ at finite source-drain voltage. This occurs for $V_\mathrm{sd}$ in the range between 0.4~mV and 1.1~mV. We observe efficient cooling in these strongly negative $\frac{dG_\mathrm{diff}}{dV_\mathrm{g}}$ regions, as demonstrated by the measured spectra of thermal vibrations in Figs.~S3a-c.

\begin{figure}[h!]
	\label{cooling}
	\centering
	\includegraphics[scale=0.8, trim = 0cm 0cm 0cm 0cm, clip=true]{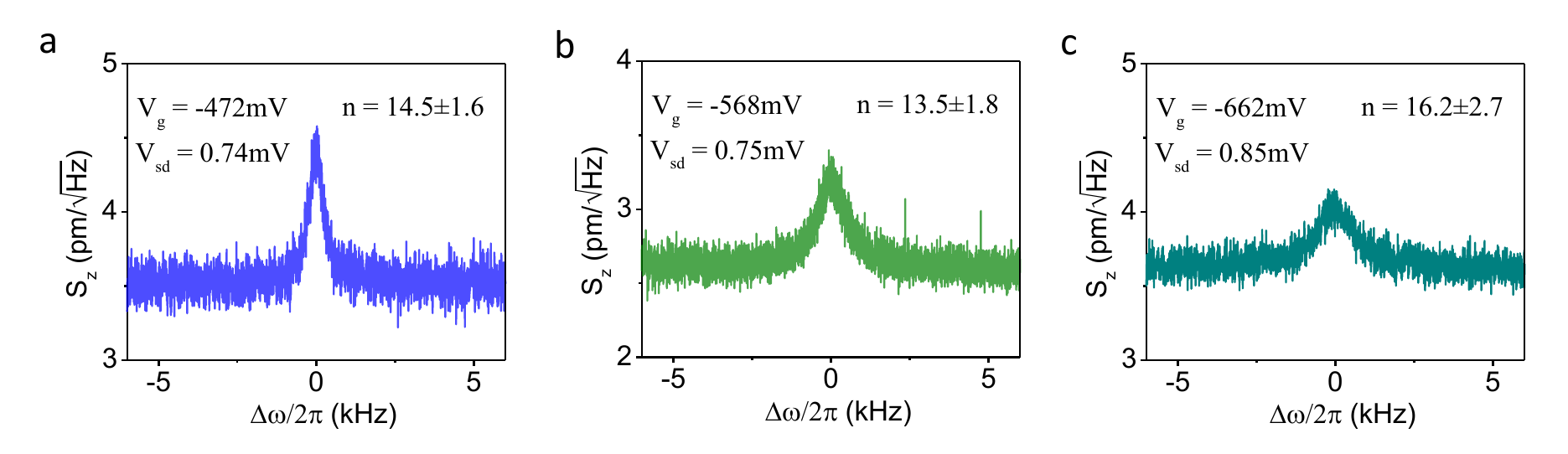}
	\caption{{\bf Cooling at different $V_g$ values.} {\bf(a-c)} Displacement power spectral density of thermal vibrations showing low occupation number in regions where $\frac{dG_\mathrm{diff}}{dV_\mathrm{g}}$ is strongly negative.}
\end{figure}

\section{Self-oscillation}

The vibrations of the nanotube in the instability region switch back and forth between thermal motion and self-oscillation, as it can be seen in the time traces of one of the two quadratures ($X$) and of the amplitude ($R$) in Fig.~S4. In these traces, the amplitude of thermal vibrations is low, whereas the amplitude in self-oscillation is high. These switches between thermal motion and self-oscillation occur randomly in time.

\begin{figure}[h!]
	\label{timedomain}
	\centering
	\includegraphics[scale=0.75, trim = 0cm 0cm 0cm 0cm, clip=true]{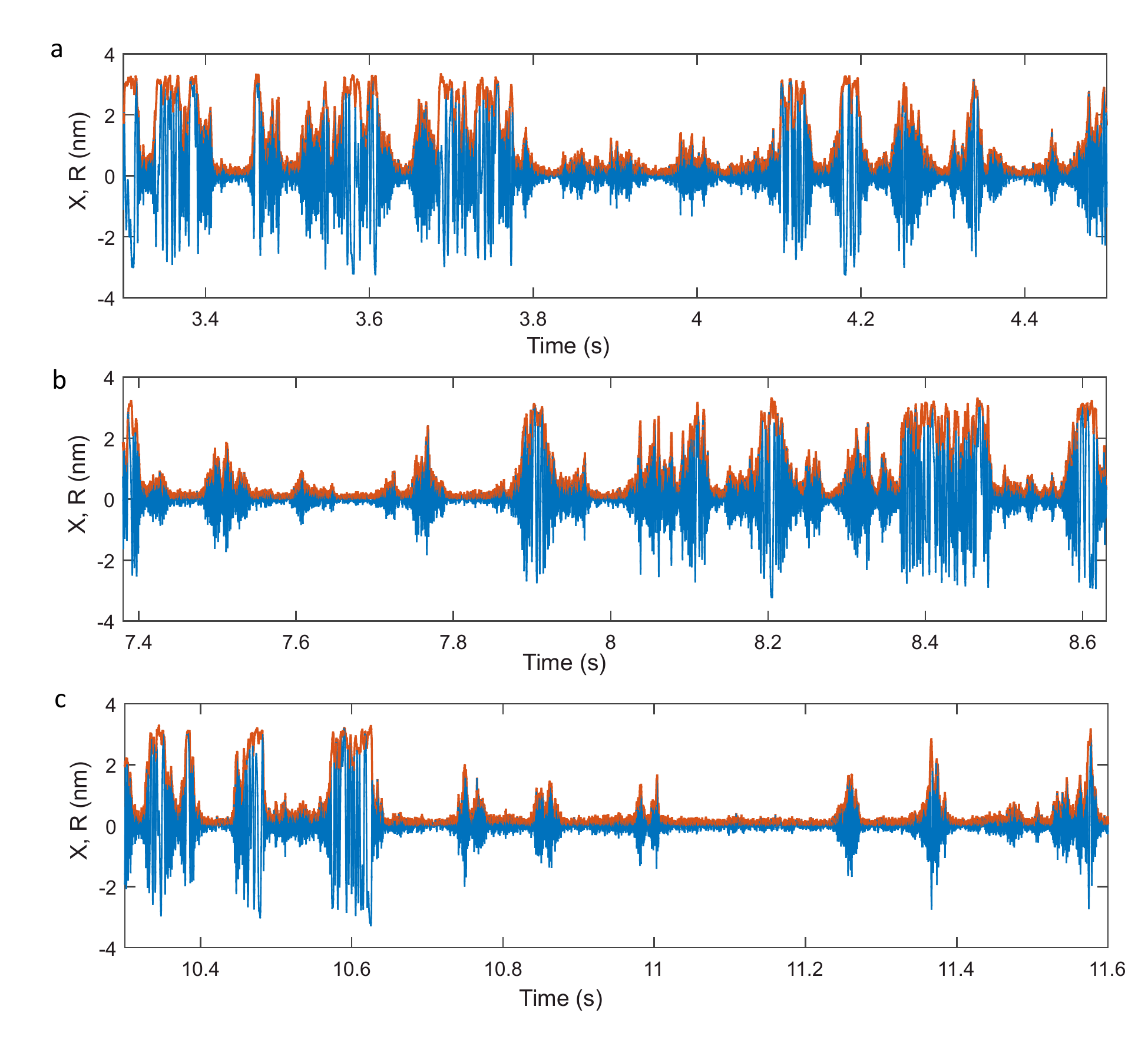}
	\caption{{\bf Time domain measurements.} {\bf(a-c)} Three different time traces of one quadrature X (blue) and corresponding amplitude R (orange) at $V_{sd}=0.25$mV for $Vg=-616$mV, plotted from the Fig. 3h in the main text.}
\end{figure}

Pure self-oscillation can also be observed without any switches to thermal vibrations. See for instance Fig.~S5. This often happens at high $V_\mathrm{sd}$ values.

\begin{figure}[h!]
	\label{timedomain}
	\centering
	\includegraphics[scale=0.75, trim = 0cm 0cm 0cm 0cm, clip=true]{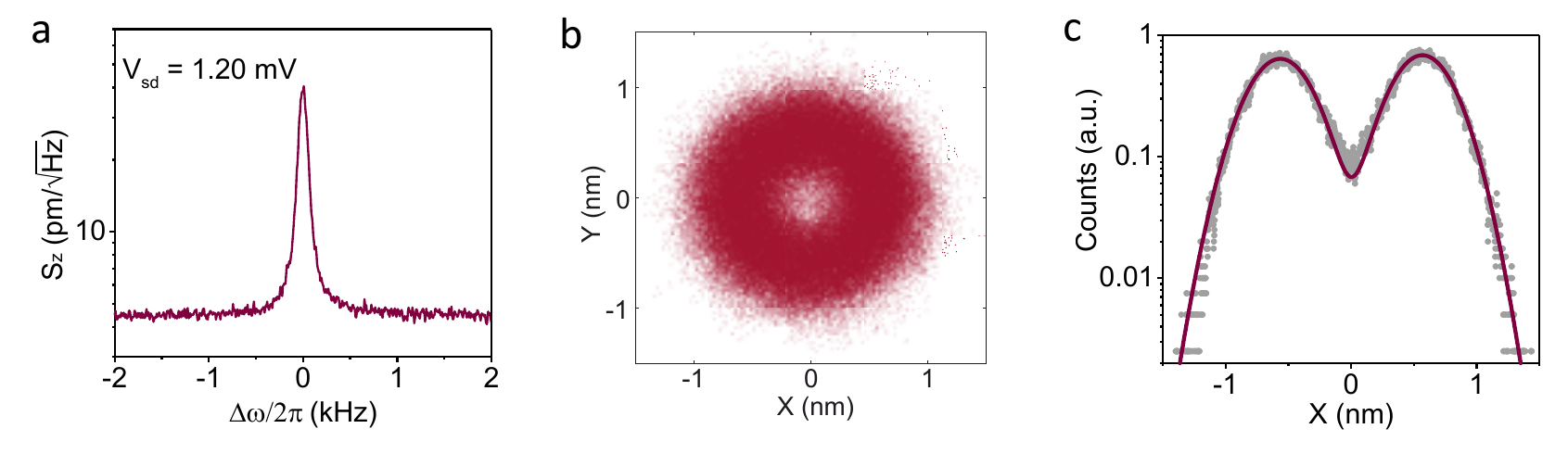}
	\caption{{\bf Pure self-oscillation.} {\bf(a)} Displacement spectral density at $V_\mathrm{sd}=1.2$~mV and $V_\mathrm{g}=-616$~mV. {\bf(b)} The phase-space of the two quadratures of the motion. {\bf(c)} Histogram associated to the phase-space in {\bf b}.}
\end{figure}

\section{Shot Noise measurement}

\begin{figure}[h!]
	\label{shotnoise}
	\centering
	\includegraphics[scale=0.75, trim = 0cm 0cm 0cm 0cm, clip=true]{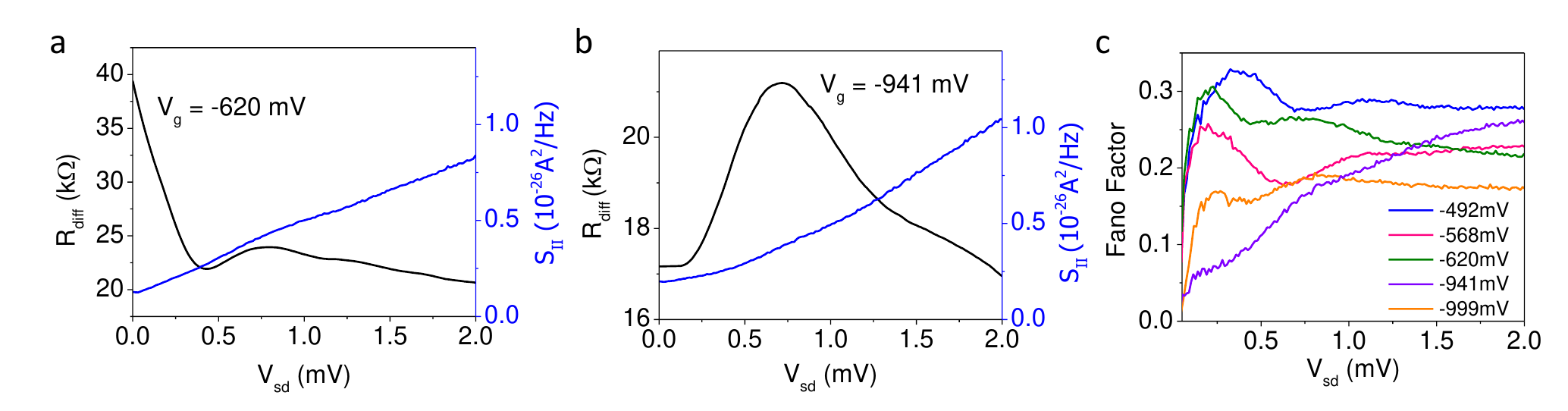}
	\caption{{\bf Shot noise measurements.} {\bf(a,b)} Differential resistance $R_\mathrm{diff}$ (grey) and current noise density $S_\mathrm{II}$ (blue) as a function of $V_\mathrm{sd}$ for two different gate voltage values. {\bf(c)} Fano factor as a function of $V_\mathrm{sd}$ for different $V_\mathrm{g}$ values.}
\end{figure}

Here we describe how we measure the shot noise of the nanotube device. The spectral density of the current noise $S_\mathrm{II}$ is transformed into spectral density of voltage noise $S_\mathrm{VV}$ through the total impedance $Z_\mathrm{tot} = (R_\mathrm{diff}^{-1}+Z_\mathrm{RLC}^{-1})^{-1}$, where $R_\mathrm{diff}$ is the nanotube differential resistance and $Z_\mathrm{RLC}$ is the effective impedance of the RLC circuit. The voltage fluctuations, which are amplified by the high-electron-mobility-transistor amplifier (HEMT), are measured at the frequency $\omega_\mathrm{RLC}= 2 \pi \cdot  1.27$~MHz over $\sim 80$kHz bandwidth. Our noise measurement contains the background contribution $S_\mathrm{II}^\mathrm{bg}$ related to the Johnson-Nyquist noise of the circuit and of the HEMT noise. This background contribution is independent of the source-drain voltage $V_\mathrm{sd}$, so that it can be quantified from the current noise measured at $V_\mathrm{sd}=0$~mV. After the substraction of this background contribution, we determine the Fano factor $F$ of the nanotube device at finite $V_\mathrm{sd}$ from the measured current noise using $F=S_\mathrm{II}(V_\mathrm{sd})/(2eI_\mathrm{sd})$, where $e$ is the electron charge, and $I_\mathrm{sd}$ is the DC current at a given source-drain bias $V_\mathrm{ds}$. The Fano factor in Fig.~S6 is smaller than one because of the suppression of the electron transmission below one and perhaps because of the electron correlation in the device.

\section{Backaction}
\subsection{Retardation time due to the circuit}
Figure~S7 shows the simplified electrical circuit used to evaluate the electrostatic and the electrothermal backactions when the retardation is given by the circuit. We consider the impedances relevant at the resonance frequency of the resonator. The nanotube with conductance $G$ is connected on the source electrode to the capacitance $C_\mathrm{RC} \simeq 60$~pF of the coaxial cable and the resistance $R_{50}=50~\Omega$ of an attenuator, which form the impedance of the circuit
\begin{equation}
Z_\mathrm{T}=\left(R_{50}^{-1}+i\omega C_\mathrm{RC}\right)^{-1}.
\label{eq:Ztot}
\end{equation}
The mechanical vibrations modulate the nanotube conductance by the amount $\delta G$. When a DC voltage $V_\mathrm{sd}$ is applied to the source electrode nanotube, the conductance modulation generates an oscillating current $\delta i_\mathrm{ac}$ at the frequency close to $\omega_0$. The current flowing through $Z_\mathrm{T}$ creates an oscillating voltage $\delta v_\mathrm{ac}$ on the source electrode, so that
\begin{eqnarray}
&& \delta v_\mathrm{ac} = - \delta i_\mathrm{ac} Z_\mathrm{T},
\label{eq:vac}\\
&& \delta i_\mathrm{ac} = \delta G V_\mathrm{sd} + G \delta v_\mathrm{ac}.
\label{eq:iac}
\end{eqnarray}
Reference~\cite{Steeneken2011} made a similar analysis as here. The difference in the two analysis comes from the fact that our device is biased with a constant voltage, while the device in Ref.~\cite{Steeneken2011} is biased with a constant current.

\begin{figure}[h!]
	\label{circuit}
	\centering
	\includegraphics[scale=1, trim = 0cm 0cm 0cm 0cm, clip=true]{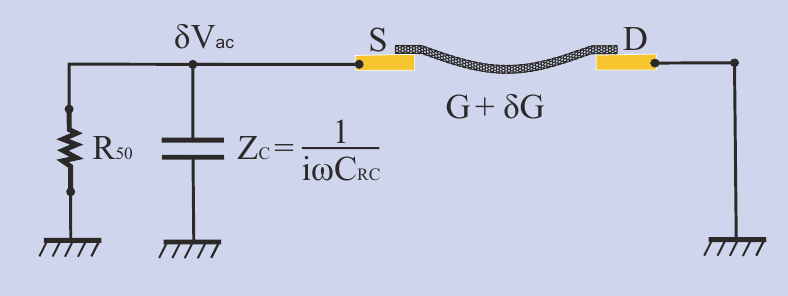}
	\caption{{\bf Simplified electrical circuit.} }
\end{figure}

The retardation time $\tau_\mathrm{RC}$ of the backaction on the vibrations is of the order of $1/\omega_0$. The retardation time is related to the delay of the modulation of $\delta v_\mathrm{ac}$ with respect to $\delta G$. We thus express $\delta v_\mathrm{ac}$ as
\begin{equation}
\delta v_\mathrm{ac} = - \frac{Z_\mathrm{T}}{1+Z_\mathrm{T}G} \delta G  V_\mathrm{sd}\simeq -R_{50}\frac{1-i\omega R_{50}C_\mathrm{RC}}{1+\omega^2 R_{50}^2C_\mathrm{RC}^2}\delta G  V_\mathrm{sd},
\label{eq:voltage}
\end{equation}
where we use $R_{50}G<<1$ in the last equality. The argument of the complex number in the numerator is $\varphi =-\mathrm{arctan}\left(\omega R_{50}C_\mathrm{RC}\right)$, so that the retardation time is
\begin{equation}
\tau_\mathrm{RC} = \frac{\mathrm{arctan}\left(\omega_0R_{50} C_\mathrm{RC}\right)}{\omega_0}.
\label{eq:tauRC}
\end{equation}
From the values of $C_\mathrm{RC}$, $R_{50}$, and $\omega_0\simeq 2\pi \cdot 92$~MHz, we get that $\omega_0R_{50} C_\mathrm{RC}=1.7$. Therefore, the retardation time $\tau_\mathrm{RC}$ of the circuit is of the order of $1/\omega_0$. The estimation $\omega_0\tau_\mathrm{RC}\sim 1$ is relevant, since this enhances cooling~\cite{Metzger2004}.

\subsection{Electrostatic backaction with the retardation due to the circuit}
As described in the last subsection, the modulation of the voltage $\delta v_\mathrm{ac}$ on the source electrode is due to the vibration-induced modulation of the conductance, when the nanotube is biased with a constant voltage. Assuming symmetric electrical contacts, the voltage modulation on the nanotube is $\delta v_\mathrm{NT}=\frac{1}{2}\delta v_\mathrm{ac}$. This results in the electrostatic force
\begin{equation}
\delta F= C_\mathrm{g}^{\prime}V_\mathrm{g}\delta v_\mathrm{NT}=-\frac{1}{2}C_\mathrm{g}^{\prime}V_\mathrm{g}R_{50}\frac{1-i\omega R_{50}C_\mathrm{RC}}{1+\omega^2 R_{50}^2C_\mathrm{RC}^2}\frac{\partial G}{\partial z} V_\mathrm{sd}\delta z.
\label{eq:electrostaticforcedelay}
\end{equation}
The real part of this backaction force leads to the shift of the spring constant, and the imaginary part to the shift of the decay rate. {\bf Using $F=-m\Delta \Gamma_\mathrm{back}\frac{dz}{dt}$ and $\frac{dz}{dt}=i\omega z$, we get}
\begin{eqnarray}
&& \Delta k_\mathrm{back} = \frac{1}{2} \left( \frac{R_\mathrm{50}} {1+(R_\mathrm{50}\omega C_\mathrm{RC})^2} \right) \frac{dG_\mathrm{diff}}{dV_\mathrm{g}} \frac{(C_\mathrm{g}^\prime V_\mathrm{g})^2 }{C_\mathrm{g}} V_\mathrm{sd},
\label{eq:omegaelectrostaticback}\\
&& \Delta \Gamma_\mathrm{back}=- \frac{1}{2m} \left( \frac{R_\mathrm{50}^2 C_\mathrm{RC}} {1+(R_\mathrm{50}\omega C_\mathrm{RC})^2} \right) \frac{dG_\mathrm{diff}}{dV_\mathrm{g}} \frac{(C_\mathrm{g}^\prime V_\mathrm{g})^2 }{C_\mathrm{g}} V_\mathrm{sd}.
\label{eq:Gammaelectrostaticback}
\end{eqnarray}
The retardation time of the backaction on the vibrations is about $1/\omega_0$.

This backaction cannot account for our data. Equation~\ref{eq:Gammaelectrostaticback} cannot account for the efficient cooling in Figs.~4a,c of the main text, since the predicted $\Delta \Gamma$ is one order of magnitude smaller than that measured in Fig.~4d of the main text.

\subsection{Electrothermal backaction with the retardation due to the circuit}
The closed loop of the backaction goes as follows. The dissipated power increases the temperature of the device. The effective thermal expansion of the device leads to the displacement of the nanotube. This displacement reacts back on the dissipated power via $\delta G =\frac{\partial G}{\partial z}\delta z$. The delay of the retardation time is $\tau_\mathrm{RC}$.

The dissipated power of the voltage-biased nanotube is $P=(G+\delta G )(V_\mathrm{sd}+\delta v_\mathrm{ac})^2$. The first-order expansion of the power reads
\begin{equation}
\delta P_\mathrm{1} = V_\mathrm{sd}^2 \delta G -2  \frac{Z_\mathrm{T}G}{1+Z_\mathrm{T}G} V_\mathrm{sd}^2 \delta G.
\label{Powerlinear}
\end{equation}
The first term of this equation leads to backaction when taking into account the thermalisation time of the device, as discussed in the next subsection. The second term results in the change of the decay rate because of the retardation of the circuit. This is what is discussed here.

The modulation of the dissipated power leads to the modulation of the mechanical tension in the nanotube. The tension modulation depends on the temperature profile along the nanotube and the electrodes, which is something hard to know precisely especially at low temperature when the electron transport is quasi-coherent~\cite{datta1996}. In what follows, we assume for simplicity that the dissipation occurs solely in the nanotube, and that temperature rises by $\delta T=\delta P\tau_\mathrm{ph}/C_\mathrm{heat}$. Here, $C_\mathrm{heat}$ is the heat capacity of the nanotube and $\tau_\mathrm{ph}$ is the thermalisation time of the nanotube. We do an additional simplification using $\tau_\mathrm{ph}\simeq L/v\simeq 0.1$~ns, where L is the nanotube length and $~v\simeq 10^4$~m/s is the phonon velocity in nanotubes~\cite{DeMartino2009}. Assuming that the thermal expansion is solely occurring in the nanotube, the nanotube expands by $\frac{\delta L}{L}=\alpha_\mathrm{TEC} \delta T$ where $\alpha_\mathrm{TEC}$ is the thermal expansion coefficient of the nanotube. Using Hook$^{\prime}$s law, the change of the mechanical tension is given by $\delta T_\mathrm{mech}=2\pi r E_\mathrm{2d}\frac{\delta L}{L}$ where $E_\mathrm{2d}=340$~N/m is the two-dimensional Young$^{\prime}$s modulus of graphene and $r$ the nanotube radius. Overall, the mechanical tension is related to the dissipated power by
\begin{equation}
\delta T_\mathrm{mech} = \frac{\alpha_\mathrm{TEC} E_\mathrm{2d} \tau_\mathrm{ph}}{C_\mathrm{heat}} 2\pi r \delta P_1.
\label{Eq:Tensionpower}
\end{equation}
We emphasize that we would get a linear relation between the tension and the power as in Eq.~\ref{Eq:Tensionpower} albeit with a different ratio $\delta T_\mathrm{mech}/\delta P_1$, if we were considering dissipation in the electrodes and/or thermal expansion of the electrodes.

The modulation of the mechanical tension generates a shift in the spring constant and in the decay rate. For this, we use the Euler$–$Bernoulli equation that reads
\begin{equation}
\rho S \frac{d^2Z}{dt^2} = -EI \frac{d^4Z}{dx^4} + \left[T_\mathrm{mech} + \frac{ES}{2L}\int^{L}_{0}\left(\frac{dZ}{dx}\right)^2dx\right]\frac{d^2Z}{dx^2}
\label{EulerBernoulli_1}
\end{equation}
where $\rho$ is the nanotube mass density, $S$ the nanotube cross-sectional area, $Z$ the displacement at the coordinate $x$ along the nanotube axis, $t$ the time, $E$ the nanotube three-dimensional Young$^{\prime}$s modulus, and $I$ the moment of inertia. We assume that the restoring force is solely given by the mechanical tension, as it is the case in our experiment, so that $EI \frac{d^4Z}{dx^4}\rightarrow 0$. We set
\begin{equation}
Z(x,t)=z_\mathrm{s}\times\phi_\mathrm{s}(x)+z_1(t)\times\phi_1(x).
\label{staticdynamicedeformation}
\end{equation}
Here, $\phi_\mathrm{s}(x)$ and $\phi_1(x)$ are the profiles of the static deformation and the measured eigenmode with $\mathrm{max}(\phi_\mathrm{s}(x))=\mathrm{max}(\phi_1(x))$=1, whereas $z_\mathrm{s}$ and $z_1(t)$ are the associated time dependent displacements. We use $\phi_\mathrm{s}(x)=\phi_1(x)=\sin(\pi x/L)$, a good approximation since the nanotube is under tensile tension. The equation of motion is obtained by multiplying the Euler$–$Bernoulli equation by $\phi_1(x)$ and integrating it along $x$. The mechanical tension is $T_\mathrm{mech}=T_\mathrm{mech}^0-\delta T_\mathrm{mech}$ where $T_\mathrm{mech}^0$ is the time-independent tension in the nanotube. The time-dependent tension creates a term proportional to $z_1$. The real part of this term induces a shift in the spring constant, and the imaginary part leads to a shift in the decay rate,
\begin{eqnarray}
&& \Delta k_\mathrm{back}=\alpha m  \frac{1}{ C_\mathrm{RC} R_\mathrm{50}} \frac{dG_\mathrm{diff}}{dV_\mathrm{g}} \frac{C_\mathrm{g}^\prime}{C_\mathrm{g}} V_\mathrm{g}z_s V_\mathrm{sd}^2,
\label{eq:omegaelectrothermalback}\\
&& \Delta \Gamma_\mathrm{back}= -\alpha\frac{dG_\mathrm{diff}}{dV_\mathrm{g}} \frac{C_\mathrm{g}^\prime}{C_\mathrm{g}} V_\mathrm{g}z_s V_\mathrm{sd}^2,
\label{eq:Gammaelectrothermalback}\\
&& \alpha=\frac{\pi^3 r}{L} \frac{\alpha_\mathrm{TEC} E_\mathrm{2d} \tau_\mathrm{ph}}{C_\mathrm{heat}} \frac{1}{m} \left( \frac{2 C_\mathrm{RC} G R_\mathrm{50}^2}{ \left(\omega C_\mathrm{RC} R_\mathrm{50}\right)^2+1}\right).
\label{eq:alphaelectrothermalback}
\end{eqnarray}
The retardation time of the backaction on the vibrations is about $1/\omega_0$, that is, $\tau \simeq 2$~ns.

We now compare the measurements of the decay rate as a function of $V_\mathrm{sd}$ in Fig.~3b and Fig.~4d with Eq.~\ref{eq:Gammaelectrothermalback} (pink lines). We estimate that the static displacement is $z_s=-0.97$~nm at $V_\mathrm{g}=-616$~mV and $z_s=-2.08$~nm at $V_\mathrm{g}=-943$~mV using $z_s=-\frac{4}{\pi}\frac{C_\mathrm{g}^\prime V_\mathrm{g}^2}{m\omega_0^2}$ from the derivation of the Euler$–$Bernoulli equation. We use $C_\mathrm{heat}=1.6\cdot10^{-22}$~J/K from Ref.~\cite{Hone2000a} where the specific heat capacity of nanotubes is $3\cdot 10^{-5}$~J/gK at 0.1~K. The only free parameter left is the thermal expansion coefficient. From the comparison between the measurements and this model, we get $\alpha_\mathrm{TEC}=9\cdot10^{-8}$~1/K. Although we did not find any report on $\alpha_\mathrm{TEC}$ for nanotubes, graphene, and graphite at such low temperatures, the order of magnitude that we get is rather realistic.

To finish this subsection, we discuss the third-order expansion of the power modulation related to Eq.~\ref{Powerlinear}, since it is relevant for the self-oscillation regime. The third-order expansion reads
\begin{equation}
\delta P_\mathrm{3} = V_\mathrm{sd}^2 \delta G^3 \left( \frac{Z_\mathrm{T}}{1+Z_\mathrm{T}G} \right)^2.
\label{Powerlinear3}
\end{equation}
Carrying out the same derivation as that described above, we obtain two additional backaction force terms, that is, a Duffing force and a nonlinear decay force of the form $z^2\frac{dz}{dt}$. Depending on the sign of $\frac{dG_\mathrm{diff}}{dV_\mathrm{g}}$, the nonlinear decay force can be negative, so that this force further increases the amplification, especially when the amplitude of motion is large. The exact derivation of the nonlinear decay force is difficult due to its renormalisation by the other nonlinear forces. The study of this nonlinear force is beyond the scope of this Letter.

\subsection{Electrothermal backaction with the retardation due to the thermalisation time of the device}
In contrast to the backaction discussed in the last subsection, this backaction arises from the modulation of the power $\delta P_\mathrm{1} = V_\mathrm{sd}^2 \delta G$ in Eq.~\ref{Powerlinear} associated to the thermalisation time of the device. The derivation of the backaction is similar to that above. The time-dependent tension that is induced by $\delta P_\mathrm{1}$ creates a force $F$ proportional to $z_1$ in the equation of motion. The shift in the decay rate is given by $\Delta \Gamma_\mathrm{back}=\frac{1}{m}\frac{\partial F}{\partial z_1}\tau_\mathrm{ph}$ when the thermalisation time $\tau_\mathrm{ph}$ is much shorter than $\omega_0$~\cite{Metzger2004}. As a result, we obtain
\begin{eqnarray}
&& \Delta \Gamma_\mathrm{back}= -\alpha\frac{dG_\mathrm{diff}}{dV_\mathrm{g}} \frac{C_\mathrm{g}^\prime}{C_\mathrm{g}} V_\mathrm{g}z_s V_\mathrm{sd}^2,
\label{eq:electrothermalrate}\\
&& \alpha=\pi^3 \frac{r}{L m} \frac{ \alpha_\mathrm{TEC} E_\mathrm{2d} \tau_\mathrm{ph}^2 }{C_\mathrm{heat}}.
\label{eq:alphaelectrothermalback2}
\end{eqnarray}

When we compare the measured $V_\mathrm{sd}$ dependence of the decay rate with this model, the agreement is satisfactory. The functional form of Eq.~\ref{eq:electrothermalrate} is the same as that in Eq.~\ref{eq:Gammaelectrothermalback} when the retardation is due to the circuit. From the comparison between the measurements and this model, we get $\alpha_\mathrm{TEC}=3\cdot10^{-9}$~1/K, which is smaller that the value obtained when the retardation is due to the circuit.

\bibliography{cooling2019}

\begin{thebibliography}{34}
\expandafter\ifx\csname natexlab\endcsname\relax\def\natexlab#1{#1}\fi
\expandafter\ifx\csname bibnamefont\endcsname\relax
  \def\bibnamefont#1{#1}\fi
\expandafter\ifx\csname bibfnamefont\endcsname\relax
  \def\bibfnamefont#1{#1}\fi
\expandafter\ifx\csname citenamefont\endcsname\relax
  \def\citenamefont#1{#1}\fi
\expandafter\ifx\csname url\endcsname\relax
  \def\url#1{\texttt{#1}}\fi
\expandafter\ifx\csname urlprefix\endcsname\relax\def\urlprefix{URL }\fi
\providecommand{\bibinfo}[2]{#2}
\providecommand{\eprint}[2][]{\url{#2}}

\bibitem[{\citenamefont{Treutlein et~al.}(2014)\citenamefont{Treutlein, Genes,
  Hammerer, Poggio, and Rabl}}]{treutlein2014}
\bibinfo{author}{\bibfnamefont{P.}~\bibnamefont{Treutlein}},
  \bibinfo{author}{\bibfnamefont{C.}~\bibnamefont{Genes}},
  \bibinfo{author}{\bibfnamefont{K.}~\bibnamefont{Hammerer}},
  \bibinfo{author}{\bibfnamefont{M.}~\bibnamefont{Poggio}}, \bibnamefont{and}
  \bibinfo{author}{\bibfnamefont{P.}~\bibnamefont{Rabl}},
  \emph{\bibinfo{title}{Hybrid Mechanical Systems}}
  (\bibinfo{publisher}{Springer, Berlin, Heidelberg}, \bibinfo{year}{2014}).

\bibitem[{\citenamefont{Aspelmeyer et~al.}(2014)\citenamefont{Aspelmeyer,
  Kippenberg, and Marquardt}}]{Aspelmeyer2014a}
\bibinfo{author}{\bibfnamefont{M.}~\bibnamefont{Aspelmeyer}},
  \bibinfo{author}{\bibfnamefont{T.~J.} \bibnamefont{Kippenberg}},
  \bibnamefont{and}
  \bibinfo{author}{\bibfnamefont{F.}~\bibnamefont{Marquardt}},
  \bibinfo{journal}{Rev. Mod. Phys.} \textbf{\bibinfo{volume}{86}},
  \bibinfo{pages}{1391} (\bibinfo{year}{2014}).

\bibitem[{\citenamefont{Knobel and Cleland}(2003)}]{Knobel2003}
\bibinfo{author}{\bibfnamefont{R.~G.} \bibnamefont{Knobel}} \bibnamefont{and}
  \bibinfo{author}{\bibfnamefont{A.~N.} \bibnamefont{Cleland}},
  \bibinfo{journal}{Nature} \textbf{\bibinfo{volume}{424}},
  \bibinfo{pages}{291} (\bibinfo{year}{2003}).

\bibitem[{\citenamefont{Woodside and McEuen}(2002)}]{Woodside2002}
\bibinfo{author}{\bibfnamefont{M.~T.} \bibnamefont{Woodside}} \bibnamefont{and}
  \bibinfo{author}{\bibfnamefont{P.~L.} \bibnamefont{McEuen}},
  \bibinfo{journal}{Science} \textbf{\bibinfo{volume}{296}},
  \bibinfo{pages}{1098} (\bibinfo{year}{2002}).

\bibitem[{\citenamefont{Lassagne et~al.}(2009)\citenamefont{Lassagne,
  Tarakanov, Kinaret, Garcia-Sanchez, and Bachtold}}]{Lassagne2009}
\bibinfo{author}{\bibfnamefont{B.}~\bibnamefont{Lassagne}},
  \bibinfo{author}{\bibfnamefont{Y.}~\bibnamefont{Tarakanov}},
  \bibinfo{author}{\bibfnamefont{J.}~\bibnamefont{Kinaret}},
  \bibinfo{author}{\bibfnamefont{D.}~\bibnamefont{Garcia-Sanchez}},
  \bibnamefont{and} \bibinfo{author}{\bibfnamefont{A.}~\bibnamefont{Bachtold}},
  \bibinfo{journal}{Science} \textbf{\bibinfo{volume}{325}},
  \bibinfo{pages}{1107} (\bibinfo{year}{2009}).

\bibitem[{\citenamefont{Steele et~al.}(2009)\citenamefont{Steele, Huttel,
  Witkamp, Poot, Meerwaldt, Kouwenhoven, and van~der Zant}}]{Steele2009}
\bibinfo{author}{\bibfnamefont{G.~A.} \bibnamefont{Steele}},
  \bibinfo{author}{\bibfnamefont{A.~K.} \bibnamefont{Huttel}},
  \bibinfo{author}{\bibfnamefont{B.}~\bibnamefont{Witkamp}},
  \bibinfo{author}{\bibfnamefont{M.}~\bibnamefont{Poot}},
  \bibinfo{author}{\bibfnamefont{H.~B.} \bibnamefont{Meerwaldt}},
  \bibinfo{author}{\bibfnamefont{L.~P.} \bibnamefont{Kouwenhoven}},
  \bibnamefont{and} \bibinfo{author}{\bibfnamefont{H.~S.~J.}
  \bibnamefont{van~der Zant}}, \bibinfo{journal}{Science}
  \textbf{\bibinfo{volume}{325}}, \bibinfo{pages}{1103} (\bibinfo{year}{2009}).

\bibitem[{\citenamefont{Benyamini et~al.}(2014)\citenamefont{Benyamini, Hamo,
  Kusminskiy, von Oppen, and Ilani}}]{Benyamini2014}
\bibinfo{author}{\bibfnamefont{A.}~\bibnamefont{Benyamini}},
  \bibinfo{author}{\bibfnamefont{A.}~\bibnamefont{Hamo}},
  \bibinfo{author}{\bibfnamefont{S.~V.} \bibnamefont{Kusminskiy}},
  \bibinfo{author}{\bibfnamefont{F.}~\bibnamefont{von Oppen}},
  \bibnamefont{and} \bibinfo{author}{\bibfnamefont{S.}~\bibnamefont{Ilani}},
  \bibinfo{journal}{Nature Physics} \textbf{\bibinfo{volume}{10}},
  \bibinfo{pages}{151} (\bibinfo{year}{2014}).

\bibitem[{\citenamefont{Ares et~al.}(2016)\citenamefont{Ares, Pei, Mavalankar,
  Mergenthaler, Warner, Briggs, and Laird}}]{Ares2016}
\bibinfo{author}{\bibfnamefont{N.}~\bibnamefont{Ares}},
  \bibinfo{author}{\bibfnamefont{T.}~\bibnamefont{Pei}},
  \bibinfo{author}{\bibfnamefont{A.}~\bibnamefont{Mavalankar}},
  \bibinfo{author}{\bibfnamefont{M.}~\bibnamefont{Mergenthaler}},
  \bibinfo{author}{\bibfnamefont{J.~H.} \bibnamefont{Warner}},
  \bibinfo{author}{\bibfnamefont{G.~A.~D.} \bibnamefont{Briggs}},
  \bibnamefont{and} \bibinfo{author}{\bibfnamefont{E.~A.} \bibnamefont{Laird}},
  \bibinfo{journal}{Phys. Rev. Lett.} \textbf{\bibinfo{volume}{117}},
  \bibinfo{pages}{170801} (\bibinfo{year}{2016}).

\bibitem[{\citenamefont{Okazaki et~al.}(2016)\citenamefont{Okazaki, Mahboob,
  Onomitsu, Sasaki, and Yamaguchi}}]{Okazaki2016}
\bibinfo{author}{\bibfnamefont{Y.}~\bibnamefont{Okazaki}},
  \bibinfo{author}{\bibfnamefont{I.}~\bibnamefont{Mahboob}},
  \bibinfo{author}{\bibfnamefont{K.}~\bibnamefont{Onomitsu}},
  \bibinfo{author}{\bibfnamefont{S.}~\bibnamefont{Sasaki}}, \bibnamefont{and}
  \bibinfo{author}{\bibfnamefont{H.}~\bibnamefont{Yamaguchi}},
  \bibinfo{journal}{Nature Communications} \textbf{\bibinfo{volume}{7}},
  \bibinfo{pages}{11132} (\bibinfo{year}{2016}).

\bibitem[{\citenamefont{Gotz et~al.}(2018)\citenamefont{Gotz, Schmid, Schupp,
  Stiller, Strunk, and Huttel}}]{Goetz2018}
\bibinfo{author}{\bibfnamefont{K.~J.~G.} \bibnamefont{Gotz}},
  \bibinfo{author}{\bibfnamefont{D.~R.} \bibnamefont{Schmid}},
  \bibinfo{author}{\bibfnamefont{F.~J.} \bibnamefont{Schupp}},
  \bibinfo{author}{\bibfnamefont{P.~L.} \bibnamefont{Stiller}},
  \bibinfo{author}{\bibfnamefont{C.}~\bibnamefont{Strunk}}, \bibnamefont{and}
  \bibinfo{author}{\bibfnamefont{A.~K.} \bibnamefont{Huttel}},
  \bibinfo{journal}{Phys. Rev. Lett.} \textbf{\bibinfo{volume}{120}},
  \bibinfo{pages}{246802} (\bibinfo{year}{2018}).

\bibitem[{\citenamefont{Singh et~al.}(2012)\citenamefont{Singh, Irfan,
  Subramanian, Solanki, Sengupta, Dubey, Kumar, Ramakrishnan, and
  Deshmukh}}]{Singh2012}
\bibinfo{author}{\bibfnamefont{V.}~\bibnamefont{Singh}},
  \bibinfo{author}{\bibfnamefont{B.}~\bibnamefont{Irfan}},
  \bibinfo{author}{\bibfnamefont{G.}~\bibnamefont{Subramanian}},
  \bibinfo{author}{\bibfnamefont{H.~S.} \bibnamefont{Solanki}},
  \bibinfo{author}{\bibfnamefont{S.}~\bibnamefont{Sengupta}},
  \bibinfo{author}{\bibfnamefont{S.}~\bibnamefont{Dubey}},
  \bibinfo{author}{\bibfnamefont{A.}~\bibnamefont{Kumar}},
  \bibinfo{author}{\bibfnamefont{S.}~\bibnamefont{Ramakrishnan}},
  \bibnamefont{and} \bibinfo{author}{\bibfnamefont{M.~M.}
  \bibnamefont{Deshmukh}}, \bibinfo{journal}{Appl. Phys. Lett.}
  \textbf{\bibinfo{volume}{100}}, \bibinfo{pages}{233103}
  (\bibinfo{year}{2012}), ISSN \bibinfo{issn}{0003-6951}.

\bibitem[{\citenamefont{Chen et~al.}(2015)\citenamefont{Chen, Deshpande,
  Koshino, Lee, Gondarenko, MacDonald, Kim, and Hone}}]{Chen2015}
\bibinfo{author}{\bibfnamefont{C.}~\bibnamefont{Chen}},
  \bibinfo{author}{\bibfnamefont{V.~V.} \bibnamefont{Deshpande}},
  \bibinfo{author}{\bibfnamefont{M.}~\bibnamefont{Koshino}},
  \bibinfo{author}{\bibfnamefont{S.}~\bibnamefont{Lee}},
  \bibinfo{author}{\bibfnamefont{A.}~\bibnamefont{Gondarenko}},
  \bibinfo{author}{\bibfnamefont{A.~H.} \bibnamefont{MacDonald}},
  \bibinfo{author}{\bibfnamefont{P.}~\bibnamefont{Kim}}, \bibnamefont{and}
  \bibinfo{author}{\bibfnamefont{J.}~\bibnamefont{Hone}},
  \bibinfo{journal}{Nature Physics} \textbf{\bibinfo{volume}{12}},
  \bibinfo{pages}{240} (\bibinfo{year}{2015}).

\bibitem[{\citenamefont{Clerk and Bennett}(2005)}]{Clerk2005}
\bibinfo{author}{\bibfnamefont{A.~A.} \bibnamefont{Clerk}} \bibnamefont{and}
  \bibinfo{author}{\bibfnamefont{S.}~\bibnamefont{Bennett}},
  \bibinfo{journal}{New J. Phys.} \textbf{\bibinfo{volume}{7}},
  \bibinfo{pages}{238} (\bibinfo{year}{2005}).

\bibitem[{\citenamefont{Armour et~al.}(2004)\citenamefont{Armour, Blencowe, and
  Zhang}}]{Armour2004}
\bibinfo{author}{\bibfnamefont{A.~D.} \bibnamefont{Armour}},
  \bibinfo{author}{\bibfnamefont{M.~P.} \bibnamefont{Blencowe}},
  \bibnamefont{and} \bibinfo{author}{\bibfnamefont{Y.}~\bibnamefont{Zhang}},
  \bibinfo{journal}{Phys. Rev. B} \textbf{\bibinfo{volume}{69}},
  \bibinfo{pages}{125313} (\bibinfo{year}{2004}).

\bibitem[{\citenamefont{Naik et~al.}(2006)\citenamefont{Naik, Buu, LaHaye,
  Armour, Clerk, Blencowe, and Schwab}}]{Naik2006}
\bibinfo{author}{\bibfnamefont{A.}~\bibnamefont{Naik}},
  \bibinfo{author}{\bibfnamefont{O.}~\bibnamefont{Buu}},
  \bibinfo{author}{\bibfnamefont{M.~D.} \bibnamefont{LaHaye}},
  \bibinfo{author}{\bibfnamefont{A.~D.} \bibnamefont{Armour}},
  \bibinfo{author}{\bibfnamefont{A.~A.} \bibnamefont{Clerk}},
  \bibinfo{author}{\bibfnamefont{M.~P.} \bibnamefont{Blencowe}},
  \bibnamefont{and} \bibinfo{author}{\bibfnamefont{K.~C.}
  \bibnamefont{Schwab}}, \bibinfo{journal}{Nature}
  \textbf{\bibinfo{volume}{443}}, \bibinfo{pages}{193} (\bibinfo{year}{2006}).

\bibitem[{\citenamefont{Zippilli et~al.}(2009)\citenamefont{Zippilli, Morigi,
  and Bachtold}}]{Zippilli2009}
\bibinfo{author}{\bibfnamefont{S.}~\bibnamefont{Zippilli}},
  \bibinfo{author}{\bibfnamefont{G.}~\bibnamefont{Morigi}}, \bibnamefont{and}
  \bibinfo{author}{\bibfnamefont{A.}~\bibnamefont{Bachtold}},
  \bibinfo{journal}{Phys. Rev. Lett.} \textbf{\bibinfo{volume}{102}},
  \bibinfo{pages}{096804} (\bibinfo{year}{2009}).

\bibitem[{\citenamefont{Santandrea et~al.}(2011)\citenamefont{Santandrea,
  Gorelik, Shekhter, and Jonson}}]{Santandrea2011}
\bibinfo{author}{\bibfnamefont{F.}~\bibnamefont{Santandrea}},
  \bibinfo{author}{\bibfnamefont{L.~Y.} \bibnamefont{Gorelik}},
  \bibinfo{author}{\bibfnamefont{R.~I.} \bibnamefont{Shekhter}},
  \bibnamefont{and} \bibinfo{author}{\bibfnamefont{M.}~\bibnamefont{Jonson}},
  \bibinfo{journal}{Phys. Rev. Lett.} \textbf{\bibinfo{volume}{106}},
  \bibinfo{pages}{186803} (\bibinfo{year}{2011}).

\bibitem[{\citenamefont{Stadler et~al.}(2014)\citenamefont{Stadler, Belzig, and
  Rastelli}}]{Stadler2014}
\bibinfo{author}{\bibfnamefont{P.}~\bibnamefont{Stadler}},
  \bibinfo{author}{\bibfnamefont{W.}~\bibnamefont{Belzig}}, \bibnamefont{and}
  \bibinfo{author}{\bibfnamefont{G.}~\bibnamefont{Rastelli}},
  \bibinfo{journal}{Phys. Rev. Lett.} \textbf{\bibinfo{volume}{113}},
  \bibinfo{pages}{047201} (\bibinfo{year}{2014}).

\bibitem[{\citenamefont{Arrachea et~al.}(2014)\citenamefont{Arrachea, Bode, and
  von Oppen}}]{Arrachea2014a}
\bibinfo{author}{\bibfnamefont{L.}~\bibnamefont{Arrachea}},
  \bibinfo{author}{\bibfnamefont{N.}~\bibnamefont{Bode}}, \bibnamefont{and}
  \bibinfo{author}{\bibfnamefont{F.}~\bibnamefont{von Oppen}},
  \bibinfo{journal}{Phys. Rev. B} \textbf{\bibinfo{volume}{90}},
  \bibinfo{pages}{125450} (\bibinfo{year}{2014}).

\bibitem[{\citenamefont{Stadler et~al.}(2016)\citenamefont{Stadler, Belzig, and
  Rastelli}}]{Stadler2016}
\bibinfo{author}{\bibfnamefont{P.}~\bibnamefont{Stadler}},
  \bibinfo{author}{\bibfnamefont{W.}~\bibnamefont{Belzig}}, \bibnamefont{and}
  \bibinfo{author}{\bibfnamefont{G.}~\bibnamefont{Rastelli}},
  \bibinfo{journal}{Phys. Rev. Lett.} \textbf{\bibinfo{volume}{117}},
  \bibinfo{pages}{197202} (\bibinfo{year}{2016}).

\bibitem[{\citenamefont{Laird et~al.}(2015)\citenamefont{Laird, Kuemmeth,
  Steele, Grove-Rasmussen, Nyg\aa{}rd, Flensberg, and Kouwenhoven}}]{Laird2015}
\bibinfo{author}{\bibfnamefont{E.~A.} \bibnamefont{Laird}},
  \bibinfo{author}{\bibfnamefont{F.}~\bibnamefont{Kuemmeth}},
  \bibinfo{author}{\bibfnamefont{G.~A.} \bibnamefont{Steele}},
  \bibinfo{author}{\bibfnamefont{K.}~\bibnamefont{Grove-Rasmussen}},
  \bibinfo{author}{\bibfnamefont{J.}~\bibnamefont{Nyg\aa{}rd}},
  \bibinfo{author}{\bibfnamefont{K.}~\bibnamefont{Flensberg}},
  \bibnamefont{and} \bibinfo{author}{\bibfnamefont{L.~P.}
  \bibnamefont{Kouwenhoven}}, \bibinfo{journal}{Rev. Mod. Phys.}
  \textbf{\bibinfo{volume}{87}}, \bibinfo{pages}{703} (\bibinfo{year}{2015}).

\bibitem[{\citenamefont{Hamo et~al.}(2016)\citenamefont{Hamo, Benyamini,
  Shapir, Khivrich, Waissman, Kaasbjerg, Oreg, von Oppen, and
  Ilani}}]{Hamo2016}
\bibinfo{author}{\bibfnamefont{A.}~\bibnamefont{Hamo}},
  \bibinfo{author}{\bibfnamefont{A.}~\bibnamefont{Benyamini}},
  \bibinfo{author}{\bibfnamefont{I.}~\bibnamefont{Shapir}},
  \bibinfo{author}{\bibfnamefont{I.}~\bibnamefont{Khivrich}},
  \bibinfo{author}{\bibfnamefont{J.}~\bibnamefont{Waissman}},
  \bibinfo{author}{\bibfnamefont{K.}~\bibnamefont{Kaasbjerg}},
  \bibinfo{author}{\bibfnamefont{Y.}~\bibnamefont{Oreg}},
  \bibinfo{author}{\bibfnamefont{F.}~\bibnamefont{von Oppen}},
  \bibnamefont{and} \bibinfo{author}{\bibfnamefont{S.}~\bibnamefont{Ilani}},
  \bibinfo{journal}{Nature} \textbf{\bibinfo{volume}{535}},
  \bibinfo{pages}{395} (\bibinfo{year}{2016}).

\bibitem[{\citenamefont{Deshpande and Bockrath}(2008)}]{Deshpande2008}
\bibinfo{author}{\bibfnamefont{V.~V.} \bibnamefont{Deshpande}}
  \bibnamefont{and} \bibinfo{author}{\bibfnamefont{M.}~\bibnamefont{Bockrath}},
  \bibinfo{journal}{Nature Physics} \textbf{\bibinfo{volume}{4}},
  \bibinfo{pages}{314} (\bibinfo{year}{2008}).

\bibitem[{\citenamefont{Moser et~al.}(2014)\citenamefont{Moser, Eichler,
  Guttinger, Dykman, and Bachtold}}]{Moser2014}
\bibinfo{author}{\bibfnamefont{J.}~\bibnamefont{Moser}},
  \bibinfo{author}{\bibfnamefont{A.}~\bibnamefont{Eichler}},
  \bibinfo{author}{\bibfnamefont{J.}~\bibnamefont{Guttinger}},
  \bibinfo{author}{\bibfnamefont{M.~I.} \bibnamefont{Dykman}},
  \bibnamefont{and} \bibinfo{author}{\bibfnamefont{A.}~\bibnamefont{Bachtold}},
  \bibinfo{journal}{Nature Nanotechnology} \textbf{\bibinfo{volume}{9}},
  \bibinfo{pages}{1007} (\bibinfo{year}{2014}).

\bibitem[{\citenamefont{Huttel et~al.}(2009)\citenamefont{Huttel, Steele,
  Witkamp, Poot, Kouwenhoven, and van~der Zant}}]{Huttel2009}
\bibinfo{author}{\bibfnamefont{A.~K.} \bibnamefont{Huttel}},
  \bibinfo{author}{\bibfnamefont{G.~A.} \bibnamefont{Steele}},
  \bibinfo{author}{\bibfnamefont{B.}~\bibnamefont{Witkamp}},
  \bibinfo{author}{\bibfnamefont{M.}~\bibnamefont{Poot}},
  \bibinfo{author}{\bibfnamefont{L.~P.} \bibnamefont{Kouwenhoven}},
  \bibnamefont{and} \bibinfo{author}{\bibfnamefont{H.~S.~J.}
  \bibnamefont{van~der Zant}}, \bibinfo{journal}{Nano Lett.}
  \textbf{\bibinfo{volume}{9}}, \bibinfo{pages}{2547} (\bibinfo{year}{2009}).

\bibitem[{\citenamefont{de~Bonis et~al.}(2018)\citenamefont{de~Bonis, Urgell,
  Yang, Samanta, Noury, Vergara-Cruz, Dong, Jin, and Bachtold}}]{Bonis2018}
\bibinfo{author}{\bibfnamefont{S.~L.} \bibnamefont{de~Bonis}},
  \bibinfo{author}{\bibfnamefont{C.}~\bibnamefont{Urgell}},
  \bibinfo{author}{\bibfnamefont{W.}~\bibnamefont{Yang}},
  \bibinfo{author}{\bibfnamefont{C.}~\bibnamefont{Samanta}},
  \bibinfo{author}{\bibfnamefont{A.}~\bibnamefont{Noury}},
  \bibinfo{author}{\bibfnamefont{J.}~\bibnamefont{Vergara-Cruz}},
  \bibinfo{author}{\bibfnamefont{Q.}~\bibnamefont{Dong}},
  \bibinfo{author}{\bibfnamefont{Y.}~\bibnamefont{Jin}}, \bibnamefont{and}
  \bibinfo{author}{\bibfnamefont{A.}~\bibnamefont{Bachtold}},
  \bibinfo{journal}{Nano Lett.} \textbf{\bibinfo{volume}{18}},
  \bibinfo{pages}{5324} (\bibinfo{year}{2018}), ISSN \bibinfo{issn}{1530-6984}.

\bibitem[{\citenamefont{Song et~al.}(2014)\citenamefont{Song, Oksanen, Li,
  Hakonen, and Sillanpää}}]{Song2014a}
\bibinfo{author}{\bibfnamefont{X.}~\bibnamefont{Song}},
  \bibinfo{author}{\bibfnamefont{M.}~\bibnamefont{Oksanen}},
  \bibinfo{author}{\bibfnamefont{J.}~\bibnamefont{Li}},
  \bibinfo{author}{\bibfnamefont{P.~J.} \bibnamefont{Hakonen}},
  \bibnamefont{and} \bibinfo{author}{\bibfnamefont{M.~A.}
  \bibnamefont{Sillanpää}}, \bibinfo{journal}{Phys. Rev. Lett.}
  \textbf{\bibinfo{volume}{113}}, \bibinfo{pages}{027404}
  (\bibinfo{year}{2014}).

\bibitem[{\citenamefont{Weber et~al.}(2016)\citenamefont{Weber, Guttinger,
  Noury, Vergara-Cruz, and Bachtold}}]{Weber2016}
\bibinfo{author}{\bibfnamefont{P.}~\bibnamefont{Weber}},
  \bibinfo{author}{\bibfnamefont{J.}~\bibnamefont{Guttinger}},
  \bibinfo{author}{\bibfnamefont{A.}~\bibnamefont{Noury}},
  \bibinfo{author}{\bibfnamefont{J.}~\bibnamefont{Vergara-Cruz}},
  \bibnamefont{and} \bibinfo{author}{\bibfnamefont{A.}~\bibnamefont{Bachtold}},
  \bibinfo{journal}{Nature Communications} \textbf{\bibinfo{volume}{7}},
  \bibinfo{pages}{12496} (\bibinfo{year}{2016}).

\bibitem[{\citenamefont{Steeneken et~al.}(2011)\citenamefont{Steeneken,
  Le~Phan, Goossens, Koops, Brom, van~der Avoort, and van
  Beek}}]{Steeneken2011}
\bibinfo{author}{\bibfnamefont{P.~G.} \bibnamefont{Steeneken}},
  \bibinfo{author}{\bibfnamefont{K.}~\bibnamefont{Le~Phan}},
  \bibinfo{author}{\bibfnamefont{M.~J.} \bibnamefont{Goossens}},
  \bibinfo{author}{\bibfnamefont{G.~E.~J.} \bibnamefont{Koops}},
  \bibinfo{author}{\bibfnamefont{G.~J. A.~M.} \bibnamefont{Brom}},
  \bibinfo{author}{\bibfnamefont{C.}~\bibnamefont{van~der Avoort}},
  \bibnamefont{and} \bibinfo{author}{\bibfnamefont{J.~T.~M.} \bibnamefont{van
  Beek}}, \bibinfo{journal}{Nature Physics} \textbf{\bibinfo{volume}{7}},
  \bibinfo{pages}{354} (\bibinfo{year}{2011}).

\bibitem[{\citenamefont{Barton et~al.}(2012)\citenamefont{Barton, Storch,
  Adiga, Sakakibara, Cipriany, Ilic, Wang, Ong, McEuen, Parpia
  et~al.}}]{Barton2012}
\bibinfo{author}{\bibfnamefont{R.~A.} \bibnamefont{Barton}},
  \bibinfo{author}{\bibfnamefont{I.~R.} \bibnamefont{Storch}},
  \bibinfo{author}{\bibfnamefont{V.~P.} \bibnamefont{Adiga}},
  \bibinfo{author}{\bibfnamefont{R.}~\bibnamefont{Sakakibara}},
  \bibinfo{author}{\bibfnamefont{B.~R.} \bibnamefont{Cipriany}},
  \bibinfo{author}{\bibfnamefont{B.}~\bibnamefont{Ilic}},
  \bibinfo{author}{\bibfnamefont{S.~P.} \bibnamefont{Wang}},
  \bibinfo{author}{\bibfnamefont{P.}~\bibnamefont{Ong}},
  \bibinfo{author}{\bibfnamefont{P.~L.} \bibnamefont{McEuen}},
  \bibinfo{author}{\bibfnamefont{J.~M.} \bibnamefont{Parpia}},
  \bibnamefont{et~al.}, \bibinfo{journal}{Nano Lett.}
  \textbf{\bibinfo{volume}{12}}, \bibinfo{pages}{4681} (\bibinfo{year}{2012}),
  ISSN \bibinfo{issn}{1530-6984}.

\bibitem[{\citenamefont{Metzger and Karrai}(2004)}]{Metzger2004}
\bibinfo{author}{\bibfnamefont{C.~H.} \bibnamefont{Metzger}} \bibnamefont{and}
  \bibinfo{author}{\bibfnamefont{K.}~\bibnamefont{Karrai}},
  \bibinfo{journal}{Nature} \textbf{\bibinfo{volume}{432}},
  \bibinfo{pages}{1002} (\bibinfo{year}{2004}).

\bibitem[{\citenamefont{Datta}(1996)}]{datta1996}
\bibinfo{author}{\bibfnamefont{S.}~\bibnamefont{Datta}},
  \emph{\bibinfo{title}{Electronic Transport in Mesoscopic Systems}}
  (\bibinfo{publisher}{Cambridge University Press}, \bibinfo{year}{1996}).

\bibitem[{\citenamefont{De~Martino et~al.}(2009)\citenamefont{De~Martino,
  Egger, and Gogolin}}]{DeMartino2009}
\bibinfo{author}{\bibfnamefont{A.}~\bibnamefont{De~Martino}},
  \bibinfo{author}{\bibfnamefont{R.}~\bibnamefont{Egger}}, \bibnamefont{and}
  \bibinfo{author}{\bibfnamefont{A.~O.} \bibnamefont{Gogolin}},
  \bibinfo{journal}{Phys. Rev. B} \textbf{\bibinfo{volume}{79}},
  \bibinfo{pages}{205408} (\bibinfo{year}{2009}).

\bibitem[{\citenamefont{Hone et~al.}(2000)\citenamefont{Hone, Batlogg, Benes,
  Johnson, and Fischer}}]{Hone2000a}
\bibinfo{author}{\bibfnamefont{J.}~\bibnamefont{Hone}},
  \bibinfo{author}{\bibfnamefont{B.}~\bibnamefont{Batlogg}},
  \bibinfo{author}{\bibfnamefont{Z.}~\bibnamefont{Benes}},
  \bibinfo{author}{\bibfnamefont{A.~T.} \bibnamefont{Johnson}},
  \bibnamefont{and} \bibinfo{author}{\bibfnamefont{J.~E.}
  \bibnamefont{Fischer}}, \bibinfo{journal}{Science}
  \textbf{\bibinfo{volume}{289}}, \bibinfo{pages}{1730} (\bibinfo{year}{2000}).

\end{thebibliography}

\end{document}